\documentclass[compsoc,conference,a4paper,10pt,times]{IEEEtran}
\IEEEoverridecommandlockouts

\usepackage{cite}
\usepackage{amsmath,amssymb,amsfonts}
\usepackage{algorithmic}
\usepackage{graphicx}
\usepackage{textcomp}
\usepackage{xcolor}

\usepackage{xurl}
\usepackage{booktabs}
\usepackage{svg}
\usepackage{pifont}
\usepackage{adjustbox}
\usepackage{float}
\usepackage[colorlinks=true,urlcolor=black]{hyperref}
\usepackage[capitalise]{cleveref}
\usepackage{multirow}

\usepackage[acronym]{glossaries}
\glsdisablehyper

 \usepackage{comment}

 \usepackage{csquotes}

\makeatletter
\renewcommand\thesection{\arabic{section}}
\renewcommand\thesubsection{\thesection.\arabic{subsection}}
\renewcommand\thesubsubsection{\thesubsection.\arabic{subsubsection}}
\def\@seccntformat#1{\csname the#1\endcsname.\quad}
\makeatother

\usepackage{titlesec}
\titleformat{\section}{\bfseries\large}{\thesection.}{0.5em}{}
\titleformat{\subsection}{\bfseries\large}{\thesubsection.}{0.5em}{}
\titleformat{\subsubsection}[runin]
  {\bfseries\normalsize}
  {\thesubsubsection.} 
  {0.5em}
  {}
  [{\unskip.\enspace}]

\titlespacing*{\section}{0pt}{2.0ex plus 1ex minus 0.5ex}{2.0ex plus 0.5ex}
\titlespacing*{\subsection}{0pt}{1.8ex plus 0.5ex minus 0.5ex}{1.5ex plus 0.3ex}
\titlespacing*{\subsubsection}{0pt}{1.5ex plus 0.5ex minus 0.2ex}{1.2ex}

\usepackage[font=footnotesize,labelfont=normalfont]{caption}

\title{Measuring Healthcare Data Leaks and Security Flaws at Internet Scale}

\makeatletter
\def\@maketitle{%
  \newpage
  \null
  \vskip 1em%
  \begin{center}%
    {\bfseries\Large \@title \par}%
    \vskip 1.5em%
    {\normalsize
      \lineskip .5em%
      \begin{tabular}[t]{c}%
        \@author
      \end{tabular}\par}%
  \end{center}%
  \par
  \vskip 1.5em}
\makeatother

\author{
  \IEEEauthorblockN{Nico Brüggemann\IEEEauthorrefmark{1}, Lukas Schmidt\IEEEauthorrefmark{2}\IEEEauthorrefmark{3}, Marvin Dölzer\IEEEauthorrefmark{1}, Marius Brockhoff\IEEEauthorrefmark{1}, Fabian Ising\IEEEauthorrefmark{1}, \\ Christoph Saatjohann\IEEEauthorrefmark{2}\IEEEauthorrefmark{3}, Sebastian Schinzel\IEEEauthorrefmark{1}\IEEEauthorrefmark{2}\IEEEauthorrefmark{3}}
  \IEEEauthorblockA{%
    \IEEEauthorrefmark{1}\textit{Fraunhofer SIT and National Research Center for Applied Cybersecurity ATHENE, Steinfurt, Germany} \\
    \textit{\{nico.brueggemann, marvin.doelzer, marius.brockhoff, fabian.ising\}@sit.fraunhofer.de} \\
    \IEEEauthorrefmark{2}\textit{FH Münster University of Applied Sciences, Steinfurt, Germany} \\
    \textit{\{lukas-schmidt, christoph.saatjohann, schinzel\}@fh-muenster.de} \\
    \IEEEauthorrefmark{3}\textit{Graduate School for Applied Research in North Rhine-Westphalia (Graduate School NRW), Bochum, Germany}
  }
}

\begin{document}
\IEEEoverridecommandlockouts
\IEEEpubid{\parbox{\columnwidth}{\copyright 2026, Nico Brüggemann. Under license to IEEE.\\DOI 10.1109/EuroSP68448.2026.00069~\hfill}
\hspace{\columnsep}\makebox[\columnwidth]{ }}

\maketitle

\IEEEpubidadjcol

\newcommand{\TotalEndpointsAuthFlaws}{2,841}

\newcommand{\HoneyUniqueIPs}{1,283}
\newcommand{\HoneyAllDICOMRequests}{3,425}
\newcommand{\HoneyDICOMEcho}{87.36}
\newcommand{\HoneyDICOMFind}{12.35}
\newcommand{\HoneyDICOMMax}{22}
\newcommand{\HoneyDICOMMin}{4}

\newcommand{\HoneyDuration}{9}

\newcommand{\HoneyIPsMulti}{754}
\newcommand{\HoneyIPsMultiPer}{58.77}
\newcommand{\HoneyIPsSingle}{529}
\newcommand{\HoneyIPsSinglePer}{41.23}

\newcommand{\AllIps}{224,634,689}
\newcommand{\AllDicomIPfour}{18,804,260}
\newcommand{\AllDicomIPsix}{129,242}
\newcommand{\AllHLsevenIPfour}{35,641,547}
\newcommand{\AllHLsevenIPsix}{336,836}
\newcommand{\AllFHIRIPfour}{166,000,704}
\newcommand{\AllFHIRIPsix}{3,722,100}

\newcommand{\AllDICOMendpoints}{9,141} %
\newcommand{\AllDICOMassociation}{4,562} %
\newcommand{\AllDICOMCEcho}{4,500} %
\newcommand{\AllDICOMhosts}{8,169} %
\newcommand{\AllDICOMIPvfour}{9,121} %
\newcommand{\AllDICOMIPvsix}{20} %

\newcommand{\AllDICOMendpointsPlain}{8,731} %
\newcommand{\AllDICOMassociationPlain}{4,496} %
\newcommand{\AllDICOMCEchoPlain}{4,434} %
\newcommand{\AllDICOMCFindPlain}{1,899} %

\newcommand{\AllDICOMendpointsTLS}{410} %
\newcommand{\AllDICOMassociationTLS}{66} %
\newcommand{\AllDICOMCEchoTLS}{66} %
\newcommand{\AllDICOMCFindTLS}{4} %

\newcommand{\AllDICOMCFindSupport}{1,903} %
\newcommand{\AllDICOMCFindData}{1,780} %
\newcommand{\AllDICOMCFindFraction}{93.54} %

\newcommand{\CFindFailedCalledAETitle}{3,355} %
\newcommand{\CFindFailedCallingAETitle}{422} %
\newcommand{\CFindFailedAETitleTotal}{3,777} %

\newcommand{\DICOMLDAP}{1,702} %
\newcommand{\DICOMLDAPtitleleak}{170} %

\newcommand{\AllHLsevenwithoutauth}{903}
\newcommand{\NoncompliantMSA}{3}

\newcommand{\AllHLseven}{917}
\newcommand{\UniqueHLSeven}{897}
\newcommand{\MSAarHLseven}{639}
\newcommand{\MSAaaHLseven}{205}
\newcommand{\MSAaeHLseven}{70}
\newcommand{\HLsevenVersion}{855}
\newcommand{\HLsevenVOther}{54}
\newcommand{\HLsevenAllPlain}{915}
\newcommand{\HLsevenAllTLS}{2}
\newcommand{\MSAaaHLsevenPlain}{204}
\newcommand{\MSAaaHLsevenTLS}{1}
\newcommand{\MSAaaHLsevenTotal}{205}
\newcommand{\MSAaeHLsevenPlain}{69}
\newcommand{\MSAaeHLsevenTLS}{1}
\newcommand{\MSAaeHLsevenTotal}{70}
\newcommand{\MSAarHLsevenPlain}{639}
\newcommand{\MSAarHLsevenTLS}{0}
\newcommand{\MSAarHLsevenTotal}{639}
\newcommand{\MSAcaHLsevenPlain}{0}
\newcommand{\MSAcaHLsevenTLS}{0}
\newcommand{\MSAcaHLsevenTotal}{0}
\newcommand{\MSAceHLsevenPlain}{0}
\newcommand{\MSAceHLsevenTLS}{0}
\newcommand{\MSAceHLsevenTotal}{0}
\newcommand{\MSAcrHLsevenPlain}{0}
\newcommand{\MSAcrHLsevenTLS}{0}
\newcommand{\MSAcrHLsevenTotal}{0}
\newcommand{\HLsevenfoundIPfour}{914}
\newcommand{\HLsevenfoundIPsix}{3}
\newcommand{\CodeAR}{69.68\%}
\newcommand{\CodeAA}{22.36\%}
\newcommand{\CodeAE}{7.63\%}
\newcommand{\CodeCA}{0.00\%}
\newcommand{\CodeSUCCESS}{22.36\%}
\newcommand{\CodeERROR}{77.32\%}
\newcommand{\MessageNotProcessedCount}{677}
\newcommand{\MessageProcessedCount}{30}
\newcommand{\MessageUnknownCount}{181}
\newcommand{\MessageEmptyCount}{177}
\newcommand{\MessageAppIssueCount}{14}

\newcommand{\AAEmptyCount}{172}
\newcommand{\AearProcessedCount}{1}

\newcommand{\AllFHIRHostlistsEndpoints}{26,536}

\newcommand{\AllFHIRHosts}{506}
\newcommand{\AllFHIR}{1,477}
\newcommand{\FHIRCapa}{460}
\newcommand{\FHIRConformance}{12}
\newcommand{\FHIRUniqueHosts}{408}

\newcommand{\AllFHIRIPvsix}{1}
\newcommand{\FHIRVersionFourOOne}{348}
\newcommand{\FHIRVersionFive}{27}
\newcommand{\FHIRAllHAPIServer}{16.00\%}

\newcommand{\FHIRanyPlain}{852}
\newcommand{\FHIRanyTLS}{625}
\newcommand{\FHIRanyany}{1,477}
\newcommand{\FHIRCapStatPlain}{314}
\newcommand{\FHIRCapStatTLS}{146}
\newcommand{\FHIRCapStatAny}{460}
\newcommand{\FHIROpOutPlain}{535}
\newcommand{\FHIROpOutTLS}{470}
\newcommand{\FHIROpOutAny}{1,005}
\newcommand{\FHIRConfStatPlain}{3}
\newcommand{\FHIRConfStatTLS}{9}
\newcommand{\FHIRConfStatAny}{12}
\newcommand{\FHIRHostsOnlyOpOut}{80}
\newcommand{\FHIRHostsOpOut}{112}

\newcommand{\FHIRPatientTotalCount}{2,124,518}
\newcommand{\FHIRPatientHostsCount}{75}
\newcommand{\FHIRPatientEndpointsCount}{242}
\newcommand{\FHIRPatientHTTPCodeOK}{225}
\newcommand{\FHIRPatientHTTPCodeUnauthorized}{280}
\newcommand{\FHIRPatientHTTPCodeNotFound}{766}
\newcommand{\FHIRPatientHTTPCodeInternalServerError}{2}
\newcommand{\FHIRPatientHTTPCodeOther}{195}
\newcommand{\FHIROrgaHTTPCodeOK}{228}
\newcommand{\FHIROrgaHTTPCodeUnauthorized}{273}
\newcommand{\FHIROrgaHTTPCodeNotFound}{778}
\newcommand{\FHIROrgaHTTPCodeInternalServerError}{4}
\newcommand{\FHIROrgaHTTPCodeOther}{186}

\newcommand{\AllServerCVE}{1,373}
\newcommand{\DICOMServerCVE}{1,333}
\newcommand{\FHIRServerCVE}{40}
\newcommand{\CVETwentyTwentyFourFiveOneOneThree}{40}
\newcommand{\CVETwentyTwoTwoOneOneNine}{397}
\newcommand{\DICOMDCMTKCount}{1,122}
\newcommand{\DICOMJDTCount}{454}
\newcommand{\DICOMOsirixCount}{337}

\newcommand{\MultidicomTotal}{8169}
\newcommand{\MultifhirTotal}{408}
\newcommand{\MultihlsevnTotal}{897}
\newcommand{\MultitotalUniqueIPs}{8984}
\newcommand{\MultiallThreeProtocols}{1}
\newcommand{\MultidicomFhirOnly}{5}
\newcommand{\MultidicomHlsevnOnly}{483}
\newcommand{\MultifhirHlsevnOnly}{0}
\newcommand{\MultidicomOnly}{7680}
\newcommand{\MultifhirOnly}{402}
\newcommand{\MultihlsevnOnly}{413}
\newcommand{\MultioverlapPercentage}{0.0}
\newcommand{\MultiHLsvenPercentage}{53.96}

\newcommand{\FHIRTLSOneZero}{0 }
\newcommand{\FHIRTLSOneOne}{0 }
\newcommand{\FHIRTLSOneTwo}{169 }
\newcommand{\FHIRTLSOneThree}{0 }

\newcommand{\HLSevenTLSOneZero}{0 }
\newcommand{\HLSevenTLSOneOne}{0 }
\newcommand{\HLSevenTLSOneTwo}{2 }
\newcommand{\HLSevenTLSOneThree}{0 }

\newcommand{\DICOMTLSOneZero}{88 }
\newcommand{\DICOMTLSOneOne}{23 }
\newcommand{\DICOMTLSOneTwo}{299 }
\newcommand{\DICOMTLSOneThree}{0 }

\newcommand{\FHIRTLSOneZeroPercent}{0}
\newcommand{\FHIRTLSOneOnePercent}{0}
\newcommand{\FHIRTLSOneTwoPercent}{100}
\newcommand{\FHIRTLSOneThreePercent}{0}

\newcommand{\HLSevenTLSOneZeroPercent}{0}
\newcommand{\HLSevenTLSOneOnePercent}{0}
\newcommand{\HLSevenTLSOneTwoPercent}{100}
\newcommand{\HLSevenTLSOneThreePercent}{0}

\newcommand{\DICOMTLSOneZeroPercent}{21.5}
\newcommand{\DICOMTLSOneOnePercent}{5.6}
\newcommand{\DICOMTLSOneTwoPercent}{72.9}
\newcommand{\DICOMTLSOneThreePercent}{0}

\newcommand{\TotalTLSOneZero}{88 }
\newcommand{\TotalTLSOneOne}{23 }
\newcommand{\TotalTLSOneTwo}{470 }
\newcommand{\TotalTLSOneThree}{0 }
\newcommand{\TotalConnections}{581 }

\newcommand{\FHIRTotal}{506 }
\newcommand{\HLSevenTotal}{917 }
\newcommand{\DICOMTotal}{8,866 }

\newcommand{\FHIRTLSPercentage}{33.4}
\newcommand{\HLSevenTLSPercentage}{0.2}
\newcommand{\DICOMTLSPercentage}{4.6}

\newcommand{\OverallTLSCount}{581 }
\newcommand{\OverallTLSPercentage}{5.6}
\newcommand{\NonTLSPercentage}{94.4}

\newcommand{\FHIRSecureCiphers}{161 }
\newcommand{\FHIRSecureCiphersPercentage}{95.3}
\newcommand{\FHIRInsecureCiphers}{8 }
\newcommand{\FHIRInsecureCiphersPercentage}{4.7}

\newcommand{\HLSevenSecureCiphers}{2 }
\newcommand{\HLSevenSecureCiphersPercentage}{100}
\newcommand{\HLSevenInsecureCiphers}{0 }
\newcommand{\HLSevenInsecureCiphersPercentage}{0}

\newcommand{\DICOMSecureCiphers}{269 }
\newcommand{\DICOMSecureCiphersPercentage}{65.6}
\newcommand{\DICOMInsecureCiphers}{141 }
\newcommand{\DICOMInsecureCiphersPercentage}{34.4}

\newcommand{\FHIRSelfSignedCount}{4 }
\newcommand{\FHIRSelfSignedPercentage}{2.4}
\newcommand{\HLSevenSelfSignedCount}{0 }
\newcommand{\HLSevenSelfSignedPercentage}{0}
\newcommand{\DICOMSelfSignedCount}{95 }
\newcommand{\DICOMSelfSignedPercentage}{23.2}

\newcommand{\TotalSelfSignedCount}{99 }
\newcommand{\TotalSelfSignedPercentage}{17.0}

\newcommand{\FHIRTotalHandshakes}{169 }
\newcommand{\HLSevenTotalHandshakes}{2 }
\newcommand{\DICOMTotalHandshakes}{410 }

 \newcommand{\LanternAllFHIRHosts}{51,443}
 \newcommand{\LanternAllFHIR}{51,442}
 \newcommand{\LanternFHIRCapa}{43,450}
 \newcommand{\LanternFHIRConformance}{36}
 \newcommand{\LanternFHIRUniqueHosts}{1,464}

 \newcommand{\LanternAllFHIRIPvsix}{0}
 \newcommand{\LanternFHIRVersionFourOOne}{41,414}
 \newcommand{\LanternFHIRVersionFive}{0}
 \newcommand{\LanternFHIRAllHAPIServer}{0.00\%}

 \newcommand{\LanternFHIRanyPlain}{1}
 \newcommand{\LanternFHIRanyTLS}{51,442}
 \newcommand{\LanternFHIRanyany}{51,443}
 \newcommand{\LanternFHIRCapStatPlain}{0}
 \newcommand{\LanternFHIRCapStatTLS}{43,450}
 \newcommand{\LanternFHIRCapStatAny}{43,450}
 \newcommand{\LanternFHIROpOutPlain}{1}
 \newcommand{\LanternFHIROpOutTLS}{7,956}
 \newcommand{\LanternFHIROpOutAny}{7,957}
 \newcommand{\LanternFHIRConfStatPlain}{0}
 \newcommand{\LanternFHIRConfStatTLS}{36}
 \newcommand{\LanternFHIRConfStatAny}{36}
 \newcommand{\LanternFHIRHostsOnlyOpOut}{7,957}
 \newcommand{\LanternFHIRHostsOpOut}{7,957}

 \newcommand{\LanternFHIRPatientTotalCount}{1}
 \newcommand{\LanternFHIRPatientHostsCount}{1}
 \newcommand{\LanternFHIRPatientEndpointsCount}{2}
 \newcommand{\LanternFHIRPatientHTTPCodeOK}{96}
 \newcommand{\LanternFHIRPatientHTTPCodeUnauthorized}{48,203}
 \newcommand{\LanternFHIRPatientHTTPCodeNotFound}{145}
 \newcommand{\LanternFHIRPatientHTTPCodeInternalServerError}{1}
 \newcommand{\LanternFHIRPatientHTTPCodeOther}{2,893}
 \newcommand{\LanternFHIROrgaHTTPCodeOK}{94}
 \newcommand{\LanternFHIROrgaHTTPCodeUnauthorized}{48,217}
 \newcommand{\LanternFHIROrgaHTTPCodeNotFound}{153}
 \newcommand{\LanternFHIROrgaHTTPCodeInternalServerError}{1}
 \newcommand{\LanternFHIROrgaHTTPCodeOther}{2,876}

\newcommand{\HLSevenDisclosureDiff}{-33}
\newcommand{\HLSevenDisclosurePercent}{-3.68}
\newcommand{\HLSevenDisclosureSame}{632}
\newcommand{\HLSevenDisclosureNew}{232}
\newcommand{\HLSevenDisclosureRemoved}{265}

\newcommand{\DICOMDisclosureDiff}{-150}
\newcommand{\DICOMDisclosurePercent}{-1.64}
\newcommand{\DICOMDisclosureSame}{6,137}
\newcommand{\DICOMDisclosureNew}{2,854}
\newcommand{\DICOMDisclosureRemoved}{3,004}

\begin{abstract}
Systems that process medical data should be meticulously secured. Yet, network services in healthcare environments often fail to implement basic security measures. For example, previous studies showed that network segmentation flaws led to DICOM systems leaking millions of patient records.

In addition to DICOM, healthcare facilities rely heavily on the HL7 and FHIR protocols to transmit data. For nine months, we operated a low-interaction honeypot for medical protocols. We found it was regularly scanned for DICOM but never for HL7 or FHIR, indicating that despite their widespread use and importance for patient data security, the security of these services remains underexplored.

In this paper, we present the first large-scale study on HL7 and FHIR services and expand previous work on DICOM. Our large-scale Internet scans, covering the three major healthcare protocols across IPv4 and IPv6 address spaces, identify healthcare systems and uncover data leaks due to authentication flaws. Additionally, we scanned for deficiencies in TLS configurations of these services and known insecure healthcare software.

In total, we found \TotalEndpointsAuthFlaws ~healthcare services with authentication flaws.
\NonTLSPercentage{}\% of all exposed systems do not support transport encryption, and \AllServerCVE{} systems have known software vulnerabilities, including those with potential for system takeover and CVSS scores up to 9.8. Overall, our study reveals an alarming state of cybersecurity in healthcare deployments, for which we discuss potential reasons and countermeasures. Finally, we report on the coordinated disclosure campaign we initiated to improve the security of patient data.

\end{abstract}


\newacronym{BSI}{BSI}{Federal Office for Information Security}
\newacronym{CERT}{CERT}{Computer Emergency Response Team}
\newacronym{MITM}{MitM}{Meddler in the Middle}
\newacronym{AE}{AE}{Application Entity}
\newacronym{AET}{AET}{Application Entity Title}
\newacronym{SCU}{SCU}{Service Class User}
\newacronym{SCP}{SCP}{Service Class Provider}
\newacronym{SOGI}{SOGI}{Sexual Orientation and Gender Identity}
\newacronym{HIS}{HIS}{Hospital Information System}
\newacronym{LIS}{LIS}{Laboratory Information System}
\newacronym{RIS}{RIS}{Radiology Information System}
\newacronym{DICOM}{DICOM}{Digital Imaging and Communications in Medicine}
\newacronym{FHIR}{FHIR}{Fast Healthcare Interoperability Resource}
\newacronym{CVE}{CVE}{Common Vulnerabilities and Exposures}
\newacronym{CVSS}{CVSS}{Common Vulnerability Scoring System}
\newacronym{MLLP}{MLLP}{Minimal Lower Layer Protocol}
\newacronym{PACS}{PACS}{Picture Archiving and Communication System}

\section{Introduction}
\label{sec:introduction}

Digitalization has become ubiquitous in healthcare facilities such as hospitals and medical centers. However, the long life cycles of expensive medical devices have led to the widespread adoption of outdated network protocols. %
These protocols often fail to implement basic security features, such as authentication or encryption, and, to protect patient data, must not be exposed to the Internet.

Nonetheless, previous studies have shown that patient data was publicly accessible due to simple network segmentation flaws, e.g., on the \gls{DICOM} network servers~\cite{doi:10.2214/AJR.15.15283, McAfeeDicom, Greenbone1, Greenbone2, chile, Blackhat2023}. As a result, these studies revealed data leaks affecting millions of patients.
The disclosure of these data leaks helped contain the exposure of particularly sensitive information, such as health status, medication, or Sexual Orientation
and Gender Identity~\cite{NAP13128}, whose public dissemination poses incalculable risks for affected persons.

Next to \gls{DICOM}, which is primarily utilized for the transmission and storage of medical image data originating from X-rays or CTs, the HL7 and \gls{FHIR} protocols are of utmost relevance in healthcare environments~\cite{TorabMiandoab2023}.
HL7 is used to control the \gls{HIS}, which stores patient data, diagnoses, and medication doses. The \gls{FHIR} standard introduces a modern, web-based successor that will become increasingly relevant in the future, and major technology companies such as Apple, Amazon, Google, IBM, and Microsoft already implemented FHIR services in their technology stacks ~\cite{fhir_tec, fhir_apple, fhir_google}.

However, to date, both HL7 and \gls{FHIR} services have been overlooked by security researchers (see our honeypot experiments in~\cref{sec:background_honeypot}). 
By focusing solely on \gls{DICOM}, previous studies explored only a fraction of the attack surface of healthcare institutions, potentially missing further critical data leaks.
To overcome these limitations, we present the first large-scale study on the security of publicly accessible HL7 and FHIR services. We further supplement our study with more thorough \gls{DICOM} measurements than previously performed. For each protocol, we scanned both the IPv4 and IPv6 address spaces for data leaks, examined the use of authentication measures and transport layer security (TLS), and identified vulnerable software versions. 

Our findings present an alarming state of cybersecurity in healthcare facilities. 
We found healthcare services disclosing patient data without authentication, systems transferring patient data without using encryption, and systems suffering from critical security vulnerabilities. We leverage the collected data to enhance patient data security and inform affected healthcare facilities during a responsible disclosure process.

Summarized, the key contributions of our work are:

\begin{itemize}
    
    \item We present the results of the first, Internet-wide measurements on the security of HL7 and FHIR services and expand on the known results for \gls{DICOM} servers. %

    \item We introduce the first study on healthcare service security in both the IPv4 and IPv6 address space.

    \item We introduce a low-interaction honeypot for \gls{DICOM}, HL7, and FHIR protocols, and present the results of a \HoneyDuration~month study on healthcare protocol attacks observed in the wild.
    
    \item We identified authentication flaws in \AllDICOMCFindData ~DICOM, \AllHLsevenwithoutauth ~HL7, and \FHIRPatientEndpointsCount ~FHIR services (\TotalEndpointsAuthFlaws ~total endpoints). 
    Furthermore, we show that \NonTLSPercentage \% of endpoints transfer patient data without encryption, and \AllServerCVE{} endpoints suffer from security flaws with a CVSS score up to 9.8.
    \item We report all identified security flaws to operators as part of a responsible disclosure process.
\end{itemize}

\section{Background}
\label{sec:background}

\subsection{Data Transmission in Healthcare Facilities}
\label{sec:background_data_transfer}

The Hospital Information System (HIS), the central patient record storage, manages and streamlines the administrative, clinical, and financial operations of a hospital.
To centralize diagnostics and laboratory or radiological findings, a \gls{HIS} usually communicates with other information systems. Typical examples are the \gls{LIS} or the \gls{RIS}~\cite{1615718, 10.5858/arpa.2012-0362-RA}. 

\begin{figure}[ht]
   \centering
   \includegraphics[width=0.9\linewidth]{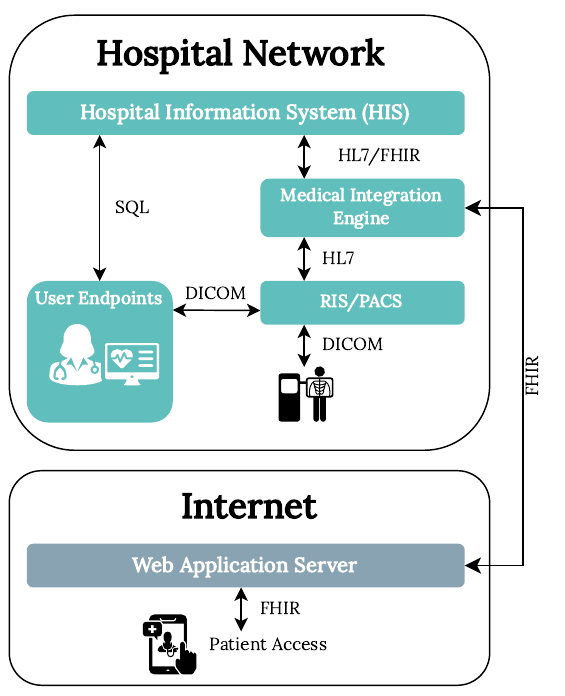}
   \caption{Exemplary transfer of data for healthcare protocols in healthcare facilities.}
   \label{fig:data_transfer}
\end{figure}

These systems are developed by multiple companies using various technologies~\cite{XU2000157}, resulting in a heterogeneous landscape and interoperability issues~\cite{Torab-Miandoab2023}. The same applies to the protocols used for data exchange between these systems, e.g., DICOM, HL7, FHIR, LIS 1/2, and EDIs. 
Moreover, proprietary protocols by the information system and device manufacturers are often used. 
However, DICOM, HL7, and more recently, FHIR are considered the most widely used protocols in healthcare facilities~\cite{FHIRSurvey, interoperabilitystandards}.
\Cref{fig:data_transfer} shows an example of communication between common information systems in a hospital.

\subsection{Healthcare Protocols}
\label{sec:background_protocols"}

\subsubsection{DICOM}
\label{sec:background_dicom}

\gls{DICOM} is a healthcare protocol used to store and transmit digital information between medical imaging devices (e.g., a CT or a MRI scanner) and other systems (e.g., a \gls{RIS} or a \gls{PACS}). DICOM was first published in 1985, and according to the Medical Imaging Technology Association (MITA), it remains one of the most widely used healthcare protocols today \cite{aboutdicom}. The DICOM standard defines a DICOM file data format and a network protocol known as the DICOM Upper Layer Protocol. 
The DICOM protocol enables communication between two DICOM systems, also referred to as \glspl{AE}~\cite{dicompart08}. AEs are identified by AE-titles and provide the usage of DICOM services, which define the functionality an AE offers to its peer.

The current version of the DICOM standard is divided into 22 parts, and unlike HL7, it defines security features. Part 15, "Security and System Management Profiles", of the DICOM standard describes support for transport encryption, digital signatures, the Secure DICOM File, which allows DICOM data to be stored in an encrypted form, and support for audit messages~\cite{dicompart15}. Part 7 describes an authentication system called User Identity Negotiation~\cite{dicompart07}. According to the standard, User Identity Negotiation supports a username and password, a Kerberos service ticket, SAML, or a JSON Web Token (JWT). However, the DICOM standard does not require vendors to implement any of these technologies, nor does it require operators to use them. In practice, this means that DICOM systems can be operated without any security measures in place. 

\subsubsection{HL7}
\label{sec:background_hl7}

HL7 (Health Level Seven) is the most widely used standard for data transmission in healthcare facilities. The organization responsible for its development, Health Level Seven International, reports that 95\% of all healthcare organizations in the US use HL7~\cite{hl7v2standard}. There are currently two versions of HL7 in use, v2 and v3. While HL7v2 was published in 1987, its direct successor, HL7v3, never achieved widespread adoption, mainly due to its complexity~\cite {HL7FHIR}. Versions 2 and 3 are fundamentally different and incompatible.

HL7v2 messages are encoded in an ASCII-based text format. A message consists of several segments, each of which consists of several fields. The fields are separated by a pipe. In most cases, data is transmitted using the \gls{MLLP} over TCP. HL7v2 has no security features, except for the User Authentication Credential Segment (UAC), introduced in version 2.6. UAC is used for user authentication and supports Kerberos and SAML. The lack of security features makes HL7v2 a vulnerable protocol, and protections must be provided entirely by the operator. HL7v2 interfaces that are exposed to the Internet can therefore be considered an immediate security risk.

\subsubsection{FHIR}
\label{sec:background_fhir}

The \gls{FHIR} standard, introduced by the HL7 international organization in 2011, aims to combine the strengths of HL7v2 and HL7v3~\cite{fhirwebsitesummary}. Although FHIR is a comparatively new protocol, its usage is steadily increasing. This is partly due to large tech companies, including Amazon, Google, IBM, Microsoft, Oracle, and Salesforce, agreeing to use FHIR to equip the healthcare sector with cloud-enabled technology~\cite{fhiriti}. The scientific literature is also increasingly focusing on FHIR, which can be interpreted as a further sign of its growing relevance~\cite{UseOfFHIR}. 

FHIR is based on modern web technologies and uses JSON and XML for data transmission via HTTPS or HTTP. Furthermore, FHIR describes itself as RESTful and is compatible with Level 2 of the REST Maturity Model~\cite{fhirwebsitererestful}.
FHIR uses resources modeled as objects in the healthcare system, such as patients, organizations, and medications. 
The current FHIR version, R5, contains 157 different FHIR resources~\cite{fhirwebsiteresource}. 

The FHIR standard itself does not prescribe any security features. However, FHIR includes a Security and Privacy Module that summarizes security topics for FHIR systems. This includes transport encryption, authorization and access control, authentication, security and privacy audit logging, digital and electronic signatures, and more~\cite{fhirwebsitesecurity}. Overall, the web-based architecture makes it much easier to secure FHIR using well-known security measures.

\section{Honeypot}
\label{sec:background_honeypot}

Previous studies \cite{doi:10.2214/AJR.15.15283, McAfeeDicom, Greenbone1, Greenbone2, chile, Blackhat2023} revealed that a considerable number of DICOM servers exposed patient data, such as X-ray images, to the Internet.
However, no studies on publicly available HL7 or FHIR endpoints have been published. 
Furthermore, it remains unknown if attackers already target these endpoints on a large scale.
To close this gap, we developed a low-interaction honeypot exposing HL7, DICOM, and FHIR endpoints. This honeypot pretends to be a healthcare system accessible on the Internet and logs any attempts to use the DICOM, HL7 or FHIR protocols on the port listed in Appendix \ref{sec:apendix_honeypot}, providing insights into the current attack landscape.

\begin{figure}%
    \centering
    \includesvg[width=1.0\linewidth]{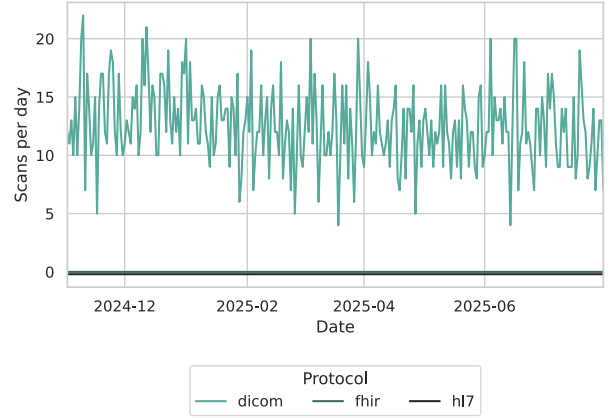}
    \caption{Number of requests registered by the honeypot over a time period of \HoneyDuration~months.}
    \label{fig:honeypot}
\end{figure}

We set up our honeypot on an independent server on the Internet, ran it from 1st November 2024 to 1st August 2025, and evaluated the collected logs. The honeypot was hosted by Netcup, a company based in Germany, and used an IP address from AS197540. As shown in~\cref {fig:honeypot}, a public DICOM server can be expected to be scanned between \HoneyDICOMMin ~and \HoneyDICOMMax ~times per day. A total of \HoneyAllDICOMRequests ~DICOM requests were registered during the analysis period. \HoneyDICOMEcho~\% of these were C-ECHO requests, and a further \HoneyDICOMFind~\% were C-FIND requests. When analyzing the source IP addresses, we saw automated large-scale scans from platforms such as Shodan or Censys. Moreover, we observed interactions from non-attributable addresses, indicating scans from potentially malicious entities. A total of \HoneyUniqueIPs ~different source IP addresses were identified. A total of \HoneyIPsMulti ~(\HoneyIPsMultiPer\%) source IP addresses scanned our honeypot multiple times, while \HoneyIPsSingle ~(\HoneyIPsSinglePer\%) were detected only once.

Interestingly, despite simulating the behavior of 14 FHIR endpoints and three HL7 services, we did not observe any interactions with them.
While this does not mean these protocols are generally ignored by attackers, our results indicate that no Internet-wide scans are currently being conducted, either by malicious parties or by researchers.

Our honeypot did not attempt to lure attackers into believing that the host in use belonged to a healthcare facility. The low-interaction honeypot exclusively mimics the endpoints of three healthcare protocols, sending a static, predefined response to requests. There is no special website or messages that could mislead the attacker into believing that this is a real system. Furthermore, the IP address does not indicate that the host is part of a healthcare facility. Despite this limitation, the honeypot is well suited to our intended use of detecting large-scale scans aimed at identifying healthcare protocols.

\section{Methodology}
\label{sec:methodology}

\subsection{Threat Model}

Patient data is highly sensitive and therefore attractive to attackers. Attackers who can access patient data can use it to blackmail patients, healthcare facilities, or sell the data directly on the darknet. Numerous such healthcare data breaches have been reported~\cite{change,medibank, wired}. 
If attackers can not only read but also modify patient data, the risks to patients may increase, e.g., to the point where changes in medication doses threaten patients' well-being.

We assume a threat model where attackers target publicly accessible healthcare systems that expose sensitive services, such as FHIR, HL7, or DICOM endpoints. %
Attackers may connect directly to the IP address of a public-facing healthcare host using standard tools, send protocol-compliant requests, and try to overcome authentication measures to extract or manipulate patient data. Furthermore, attackers may choose to perform \gls{MITM} attacks, trying to intercept network traffic and capture sensitive user data.

\begin{figure*}[ht!]
   \centering
   \includegraphics[width=1.0\linewidth]{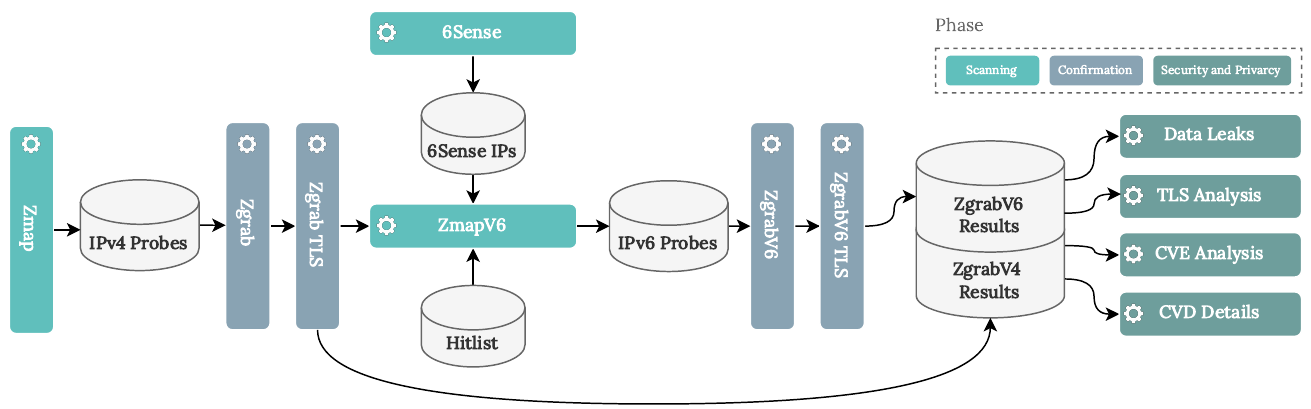}
   \caption{Overview of the IT architecture of the scanning pipeline. The pipeline is run through for all three healthcare protocols.}
   \label{fig:scan_pipeline}
\end{figure*}
\subsection{Overview}

We split up our analysis into three phases: The Scanning phase, the Confirmation phase, and the Security \& Privacy phase. During the Scanning phase, we scanned public IPv4 and IPv6 addresses on ports associated with healthcare protocols (see~\cref{tab:scan_results_ports}) to identify potential publicly exposed healthcare services.
During the Confirmation phase, we exchange protocol handshakes with these systems, ensuring that they serve the expected network protocol. After successfully identifying healthcare systems, we investigated them more thoroughly during the Security \& Privacy phase.

However, the Scanning and Confirmation phases do not run in parallel or one after the other. Rather, the two phases overlap. First, an IPv4 scan is started (scanning phase). Once this is complete, the IPv4 confirmation phase for non-TLS and TLS is started automatically. Next, the IPv6 scan (scanning phase) is performed, after which the IPv6 confirmation phase for non-TLS and TLS is started. The final step in the process is the Security \& Privacy phase. We depicted the entire scanning process in~\cref{fig:scan_pipeline}. The same scan process is run for all three healthcare protocols.

\subsection{Scanning phase}

\subsubsection{IPv4 and IPv6 scanning}
For identifying healthcare services at Internet scale, we used ZMap~\cite{durumeric2013zmap} and ZMapv6~\cite{zmapv6}, network scanners that are commonly used for Internet-wide measurements.
While scanning the whole IPv4 address range (/0) using ZMap is straightforward, it is impossible to scan the full IPv6 range due to the vast quantity of IPv6 addresses ($2^{128}$).

To scan as many systems as possible, we combined two methods to obtain lists of promising IPv6 address ranges.
First, we used 6Sense~\cite{298086}, a research project that leverages machine learning to generate and optimize IPv6 address lists for active hosts. 
We ran 6Sense on a Linux server with 512~GB of memory and two 80~GB Nvidia A100 GPUs, which allowed us, over the course of a week, to generate an address list holding 100 million IPv6 addresses. 
We combined this address list with the IPv6 addresses from the widely known IPv6 Hitlist~\cite{gasser2018clusters,zirngibl2022rustyclusters,steger2023targetacquired}, a research project that collects and shares lists of actively communicating IPv6 addresses to facilitate Internet measurements. 
We removed duplicates arising from combining the two address lists and fed our results into ZMapv6 to perform our scans. Depending on the healthcare protocol, this approach resulted in between 137 and 139 million IPv6 addresses, since a different number of ports were scanned for each protocol. The number of IPv6 addresses (n=139 million) is significantly lower than the total number of IPv4 addresses (n=3.70 billion non-reserved).

Since we scan both IPv4 and IPv6 hosts during the Scanning phase, and we need to minimize the time between the Scanning phase and the Confirmation phase to minimize the effect of dynamic IP address churn, we first performed scans and confirmations for IPv4 addresses, then for IPv6 addresses. The data collected is based on a single scan per healthcare protocol. The scans took place in October 2025.

\subsubsection{Port list}
During the Scanning phase, we scanned the Internet for hosts listening on TCP ports that are commonly used by healthcare services. 
As no comprehensive list of commonly used ports for DICOM, HL7, and FHIR services exists, we distilled one using the following multi-step approach.

Firstly, we analyzed the healthcare protocol standards and collected information on officially registered ports from the Internet Assigned Numbers Authority (IANA). For example, ports 11112, 2762, and 2761 are reserved for DICOM at IANA~\cite{iana}. 
Secondly, we investigated which standard ports are used in major software implementations, e.g., by reading the software documentation or directly examining the source code in public Git repositories. 
We assume that a significant number of administrators do not change preconfigured ports. 
Thirdly, as a supplementary method for FHIR services, we used public data from Censys.io~\cite{censys} and Shodan.io~\cite{shodan}, both popular online platforms that provide access to regularly conducted Internet scans. Both services allow filtering search results and examining the service responses to HTTP queries. Since FHIR systems may listen on a multitude of ports, we searched their scan databases for the string \texttt{application/fhir+}, which is used by most FHIR services as the Content-Type header in HTTP responses. 
In this way, we obtained a list of actively listening FHIR services and added the most commonly found ports as scan targets.

In total, we collect 31 ports for our scans. \cref{tab:scannedports} in the appendix provides an overview of the scanned ports and their origin.

\subsection{Confirmation phase}

Previous work has shown that administrators, besides using preconfigured ports, regularly choose to run protocols on non-standard ports or on ports that may be reserved for other protocols~\cite{LZR}. We therefore need to verify that services running on common ports indeed run DICOM, HL7, or FHIR~\cite{highly_responsive}. During the Confirmation phase, we connect to the previously identified candidates, send standard-conforming protocol messages to the services, and verify the responses.
For this, we developed Zgrab2~\cite{zgrab} modules that send protocol-specific messages and examine the services' responses. 

For each service, we tried to establish a connection once with TLS and once without TLS, acquiring the data required for our TLS analysis (see \cref{sec:confimation_privacy}).

\subsubsection{DICOM Confirmation}

When a DICOM client --~a \gls{SCU}~-- initiates communication with a DICOM server --~a \gls{SCP}~--, the first step is the negotiation of connection parameters in the form of a handshake. 
This process is named \textit{Association Negotiation}, and results in coordinated communication parameters between the AEs~\cite{dicompart08}, e.g., format and order of messages.

We use this process to verify the presence of DICOM services; i.e., we initiate DICOM association negotiations with potential DICOM services.
Based on the go-dicom package~\cite{godicom}, we developed a ZGrab2 module that contacts the \gls{SCP} with an association request (A-ASSOCIATE-RQ). 
We request the usage of the DICOM C-ECHO service, a service that is commonly used to verify application-level communication between AEs. 
Upon receiving the association request, the SCP evaluates the proposed parameters and sends an A-ASSOCIATE-AC response, including information about acceptance or rejection of the handshake.
In case we receive a response, regardless of acceptance status, we successfully identified an active DICOM service instance.

\subsubsection{HL7 Confirmation}
\label{sec:confirmation_hl7}
To identify HL7 systems, we developed a ZGrab2 module that sends an HL7 message over MLLP to the remote system. 
By ensuring that the HL7 message conforms to the IHE IT Infrastructure (ITI) Technical Framework~\cite{ihe}, a common set of components and interactions used by healthcare facilities, we ensured that it can be processed by most HL7 systems. 

As HL7 allows the manipulation of patient records, we send an HL7 message type that definitely does not change any patient data: the HL7 Patient Demographics Query (PDQ) message. 

To prevent the return of real patient data, we specifically filter for a made up and most likely non-existing patient with the first name \textit{Mr.} and the last name \textit{Spock}. The resulting HL7 message is shown in \cref{fig:hl7_msg} in the appendix.

Although a PDQ query may not be supported by all systems, and a search for the patient \texttt{Mr. Spock} may not return any results, we can still use this message to confirm the presence of an HL7 service. 
In either case, the remote system responds with a Message Acknowledgement (MSA), on which we mark the service as confirmed. 
An MSA contains a field (MSA-1) with an acknowledgement code that can have one of the following values: Application Accept (AA), Application Error (AE), Application Reject (AR), Commit Accept (CA), Commit Accept (CA) or Commit Reject (CR)~\cite{hl7v2_02}. The acknowledgement codes show how the remote system handled our message.

\subsubsection{FHIR Confirmation}
\label{subsubsec:MethFHIR}
The presence of a FHIR \texttt{Capability\-Statement} can be used to identify FHIR endpoints. Capability Statements are used to retrieve metadata and provide an overview of the features implemented by a FHIR endpoint. As Capability Statements contain technical information important to clients rather than sensitive information, they can be retrieved without authentication. If an error occurs while retrieving any FHIR resource, the FHIR endpoint returns an \texttt{OperationOutcome} error. This error message is also FHIR-specific and can therefore be used to identify an FHIR endpoint. However, an OperationOutcome error contains significantly less information about the system than a Capability Statement.

A major challenge in searching for FHIR systems is the freely selectable URLs of FHIR endpoints. A URL for a FHIR endpoint consists of the IP address or the domain name and the FHIR endpoint name, as well as the resource type at the end of the URL, e.g. \texttt{/metadata}, which is used to query a Capability Statement. A FHIR URL has the following structure: \texttt{http://\allowbreak\{ip/domain\}/\allowbreak\{endpoint\}/\allowbreak\{resource\_type\}}. From the ZMap scan, we know the IP addresses and the resource type (\texttt{/metadata}). However, the \texttt{\{endpoint\}} remains unknown. To find as many FHIR endpoints as possible, we collect the standard endpoints of FHIR applications. We examined several lists of FHIR applications and public FHIR test servers \cite{fhirserverlist, fhirtestserver} and extracted their endpoints. In total, we identified 13 endpoints using this method. A complete list of the endpoints used can be found in \cref{sec:FHIREndpoints}.

Each of these endpoints is queried by our ZGrab FHIR module for each IP address. We also translate IP addresses into possible hostnames, thereby avoiding problems when accessing servers that use virtual hosting. To achieve this, we extract the common name (CN) from the TLS certificates of the web servers, as this method should retrieve the names that best match the virtual host entries.
For each virtual host, 13 queries, one per endpoint, need to be sent to the server, leading to a high number of queries per server, which we want to avoid. Taking the information in the Subject Alternative Name (SAN) field into account could provide more addresses. However, this would lead to a significant increase in the number of requests per host. Our approach ensures that virtual hosts can be queried while limiting the number of queries per IP address to 26. 

The ZGrab module then searches the requested FHIR URLs for the strings \texttt{CapabilityStatement}, \texttt{Conformance} (this string is used up to FHIR version DSTU2 in a CapabilityStatement), and \texttt{OperationOutcome}. If at least one of these strings is present, an FHIR endpoint has been found.

There are also some companies that offer FHIR endpoints as a service for customers or patients. They use a UUID or a complex random string for the \texttt{\{endpoint\}} part of the URL, enabling them to assign each customer their own endpoint. Therefore, it is not feasible to detect these endpoints with our scans. Nevertheless, to account for FHIR endpoints with arbitrary URLs, we have included an analysis of the Lantern dataset, which is a collection of FHIR endpoints, in Appendix \ref{sec:Lanter}. This dataset provides a collection of FHIR endpoints from developers of healthcare IT products.

\subsection{Security and Privacy phase}
\label{sec:confimation_privacy}

During the security and privacy phase, we thoroughly investigated potential data leaks, TLS flaws, and software vulnerabilities across the identified healthcare services.
To do so, we enriched our scan results with additional data from Censys.io~\cite{censys15}. This includes server geolocation, abuse information for the disclosure process, and information about additional ports the servers listen on. 

\subsubsection{Identifying DICOM Data Leaks}
While public exposure of a DICOM service does not directly imply a data leak, such a leak occurs as soon as the service provides access to patient data without authentication. 
Previous studies have shown that for authentication~\cite{Greenbone1, Greenbone2, Blackhat2023}, DICOM services often rely solely on the AE-titles transmitted during the Association Negotiation. In these cases, whenever a valid AE-title is recognized by the server, it provides access to patient data. Even more concerning, DICOM services may choose to accept arbitrary AE-titles, which corresponds to the absence of any authentication. %

We identify data leaks by attempting to establish an association using an arbitrary AE-title and, upon successful authentication, use the DICOM C-FIND service to test access to patient data. 
C-FIND requests allow clients to query a remote server for patient information. On success, the server returns the requested patient records with the corresponding status code. 
Due to its scriptable Python API, we use pynetdicom~\cite{pynetdicom} to perform the described analyses. 
To ensure that our study meets the highest possible ethical standards, we process as little data as possible when verifying access to patient data. We only request the first available patient record and only evaluate the status code of the response for verification.
We do not process or store any further patient data, minimizing the data required for leak verification to the absolute minimum. 

Similarly, while brute-forcing AE-titles based on a list of commonly chosen values may certainly be feasible and interesting from a security perspective, we considered this approach problematic for ethical and legal reasons. 
Therefore, we decided to omit brute-forcing and instead use a side-channel to study chosen AE-titles. Previous work on LDAP servers has indicated that healthcare services may leak AE-titles via LDAP instances running on the same server~\cite{kaspereit}. Therefore, we scan healthcare servers for LDAP instances running on the same IP address, and evaluate the AE-titles they leak. %

\subsubsection{Identifying HL7 Data Leaks}

Detecting HL7 data leaks is significantly more complex than simply identifying HL7 endpoints. A data leak requires data to be readable. However, HL7 is not primarily designed for reading data. It is much more designed for pushing messages, i.e., writing and updating data records. Nevertheless, it is possible to detect a data leak. To achieve this, we evaluate the response to the PDQ message described in \cref{sec:confirmation_hl7} that we send, even though the PDQ function is not supported by most endpoints.

The received MSA message can be divided into two cases. Either the HL7 endpoint processed our PDQ message, or an error occurred during processing. If the PDQ message was processed, the MSA message will contain the value \texttt{AA} or \texttt{CA} in the MSA-1 field, and the MSA-3 field will contain text that can be selected by the healthcare facility operator. In this case, we assume that we can perform any PDQ query to search for specific patient data, which could result in a data leak. If an error occurs during processing of the PDQ message, however, the MSA-1 field contains the value \texttt{AE}, \texttt{AR}, \texttt{CA}, or \texttt{CR}. Additionally, an error segment will be added to the HL7 message. The ERR-3 field in the error segment contains an error code, while the ERR-8 field contains free text that can be selected by the operator.

The MSA message can also be used to determine whether the HL7 endpoint uses the UAC authentication. If the server supports UAC but authentication information is missing or incorrect, the ERR-7 field must indicate an authentication error~\cite{hl7v2standard}. 

\subsubsection{Identifying FHIR Data Leaks} 
\label{sec:IdentifyingDataLeaksFHIR}
Unlike DICOM and HL7, many FHIR services are intended for patient access and information sharing, and therefore need to be exposed to the Internet. To further explore security and privacy issues, we request multiple resources from the FHIR endpoints. The \texttt{CapabilityStatement} resource contains information about the FHIR version in use, the implementation name and version, and the supported authentication methods. %

A FHIR data leak is present when the \texttt{/Patient} resource can be accessed without authentication. However, patient data would be transmitted directly when the \texttt{/Patient} resource is accessed from a service without required authentication. In line with our ethical considerations (see \cref{sec:ethics}), we have therefore decided to use the summary function of the FHIR standard~\cite{fhirwebsitesummary}. This function can count the number of entries (IDs) in a resource without displaying or transmitting the actual data. The return value is the number of IDs in the database. 

To check for an FHIR leak, we query the following URL: \texttt{BASE\_URL/Patient?\_summary=count}. This enables us to detect and quantify data leaks by counting patient IDs. Technically, it is possible to use an authorization interceptor function at application level, as implemented by the HAPI FHIR server\cite{AuthorizationInterceptor}, to prevent access to all patients while allowing the summary function to count all patients. However, this is not a default configuration and we do not consider this a likely scenario, which is why we assume that patient data can also be retrieved when the summary function is successfully called. 

One disadvantage of this approach is that some FHIR software implementations do not support the summary function. Additionally, we analyse the FHIR resource \texttt{/Organization}, which is typically protected by authentication and does not expose any patient data. Furthermore, we analyse HTTP response codes to determine if authentication is required to access FHIR resources.

\subsubsection{TLS Evaluation}
Secure transport encryption is vital for protecting patient data, especially on Internet-exposed hosts. Previous work has shown that Internet-facing systems do not always configure TLS securely~\cite{Holz_2016,doi:10.3233/JCS-200070}. Therefore, we analyze the TLS configurations of the healthcare systems we discover. We analyze overall TLS usage, TLS versions and cipher suites, and check for self-signed TLS certificates.

We use ZGrab's TLS logs from the confirmation phase for the TLS analysis. We use the standard TLS configuration for Zgrab. This means that no minimum or maximum TLS versions are specified, and the standard cipher suites from ZCrypto are used~\cite{zcrypto}. The server's selected cipher suite can be retrieved from the TLS logs. We use information from ciphersuite.info, retrieved via its API, to determine which cipher suites are considered insecure~\cite{ciphersuite}.

\subsubsection{Vendor and CVE Analysis}

During the security and privacy phase, we evaluate the software implementation used by healthcare providers. For FHIR, the \texttt{/metadata} endpoint provides information about the software and its version, as well as the FHIR version, allowing detailed analysis. For DICOM, information from a C-ECHO can be evaluated as it contains details of the software library and version used. However, it is not always possible to draw a definitive conclusion about the DICOM server software from the library used, as many software manufacturers rely on third-party libraries. HL7 is the only protocol that does not offer any possibility to infer the software used directly. In this case, it is only possible to evaluate the HL7 version used, as this is always included in an ACK message.

We use the extracted software information to identify known vulnerabilities in the server implementations, as these often lead to data leaks or other security risks. For this, we match the software information with publicly known vulnerable versions based on the NIST National Vulnerability Database (NVD) API~\cite{NIST_API}.
We use the following procedure to determine the \gls{CVSS} values from the NVD: First, we 
attempt to obtain the \gls{CVSS} NVD value for version 3.x. If this value is not available, we use the value assigned by the CNA for version 3.x as a fallback. For older CVEs, we use the CVSS value for version 2.0 as a fallback if no corresponding value for version 3.x is available.

\subsubsection{Global Exposure}

To classify data leaks, TLS flaws, and vulnerabilities in healthcare systems more accurately, we match affected hosts' IP addresses to the respective countries. TLS flaws are considered only if there is an insecure TLS configuration. Hosts that do not use TLS are excluded from these statistics. The GeoLite2-Country Database \cite{geolite} is used to determine the geographical location.

\section{Scan Results}
\label{sec:results}

\subsection{Port Scan}

During the scanning phase, we found \AllIps{} hosts responding to a SYN request on one of the ports outlined in Table \ref{tab:scan_results_ports}. For DICOM, we identified \AllDicomIPfour{} potential IPv4 hosts and \AllDicomIPsix{} IPv6 hosts. For HL7, we found a total of \AllHLsevenIPfour{} IPv4 hosts and \AllHLsevenIPsix{} IPv6 hosts. The most significant number of potential hosts was found for FHIR, with \AllFHIRIPfour{} IPv4 hosts and \AllFHIRIPsix{} IPv6 hosts. The higher values can primarily be explained by the large number of hosts exposing ports 80 and 443 for regular web traffic.

\subsection{Healthcare Services at Internet-Scale}

\subsubsection{Overview}
During the confirmation phase, we found all three healthcare protocols on the Internet. The following values refer to the number of endpoints, which consist of an IP address (host) and a port. For FHIR, the endpoint is also included in the URL. This distinction is important because some hosts offer multiple instances simultaneously.

In total, we found \AllDICOMendpoints{} DICOM endpoints (listening on \AllDICOMhosts{} unique hosts). However, a DICOM association only works with \AllDICOMassociation{} endpoints, and \AllDICOMCEcho ~endpoints responded to a C-ECHO request.

Moreover, we found \AllHLseven ~HL7 endpoints (\UniqueHLSeven ~unique hosts). \MSAarHLseven ~endpoints responded with an \texttt{AR} code. \MSAaaHLseven ~endpoints responded with an \texttt{AA} code and \MSAaeHLseven ~others with an \texttt{AE} code.

Finally, we found a \AllFHIR{} FHIR endpoints (\FHIRUniqueHosts ~unique hosts) with \texttt{\//metadata} infos. \FHIRCapa ~of these use a Capability Statement, while \FHIRConformance{} provide a Conformance Statement. Furthermore, we found \FHIROpOutAny ~endpoints that return an OperationOutcome. An OperationOutcome is displayed when an FHIR endpoint is accessed via an incorrect URL. For example, if the correct FHIR URL is \texttt{/fhir/metadata}, but a GET request is made to \texttt{baseR4/metadata}, an OperationOutcome may occur. Since we query up to 13 endpoints, we often receive an OperationOutcome. For \FHIRHostsOnlyOpOut ~hosts, we only receive an OperationOutcome and no CapabilityStatement or ConformanceStatement. The stated reasons for this behavior range from \texttt{not supported} as a message in the returned resource to HTTP Code 401 unauthorized. \Cref{tab:scan_results_01} shows all identified endpoints, including status and TLS usage.

\subsubsection{IPv4 vs. IPv6}

By far, the majority of identified healthcare services use IPv4 addresses. But a few systems also use IPv6 addresses. We identified a total of \AllDICOMIPvsix ~DICOM hosts, \HLsevenfoundIPsix ~HL7 hosts, and \AllFHIRIPvsix ~FHIR host with an IPv6 address. The HL7 and FHIR hosts provide their services exclusively via IPv6 addresses. This evaluation is not possible for DICOM, as we do not have sufficient details about endpoints to clearly identify them across multiple IP addresses.

\begin{table*}[ht]
    \caption{Endpoints publicly exposing healthcare services via the Internet.}
    \label{tab:scan_results_01}
    \footnotesize
    \centering
    \begin{tabular}{lrrrlrrrlrrr}
    \toprule
        \multicolumn{4}{c}{\textbf{DICOM}}                                 & \multicolumn{4}{c}{\textbf{HL7}}                                   & \multicolumn{4}{c}{\textbf{FHIR}}                                   \\ \midrule
        \multicolumn{1}{c}{\textbf{Status}} & \textbf{Plain} & \textbf{TLS} & \textbf{$\sum$} & \multicolumn{1}{c}{\textbf{Status}} & \textbf{Plain} & \textbf{TLS} & \textbf{$\sum$} & \multicolumn{1}{c}{\textbf{Status}} & \textbf{Plain} & \textbf{TLS} & \textbf{$\sum$} \\ \midrule
        DICOM Support                       & \AllDICOMendpointsPlain & \AllDICOMendpointsTLS          & \AllDICOMendpoints & HL7 Support                         & \HLsevenAllPlain            & \HLsevenAllTLS            & \AllHLseven & FHIR Support                        & \FHIRanyPlain              & \FHIRanyTLS            & \FHIRanyany  \\
        Association                         & \AllDICOMassociationPlain           & \AllDICOMassociationTLS           & \AllDICOMassociation & MSA AA                              & \MSAaaHLsevenPlain            & \MSAaaHLsevenTLS           & \MSAaaHLsevenTotal & CapabilityStatement                 & \FHIRCapStatPlain              & \FHIRCapStatTLS            & \FHIRCapStatAny \\
        C-ECHO Response                     & \AllDICOMCEchoPlain           & \AllDICOMCEchoTLS          & \AllDICOMCEcho & MSA AE                              & \MSAaeHLsevenPlain             & \MSAaeHLsevenTLS              & \MSAaeHLsevenTotal & OperationOutcome                    & \FHIROpOutPlain              & \FHIROpOutTLS            &  \FHIROpOutAny\\
        C-FIND Support                              &   \AllDICOMCFindPlain            &  \AllDICOMCFindTLS   & \AllDICOMCFindSupport & MSA AR                              & \MSAarHLsevenPlain            & \MSAarHLsevenTLS            & \MSAarHLsevenTotal & ConformanceStatement                & \FHIRConfStatPlain              & \FHIRConfStatTLS            & \FHIRConfStatAny \\
                                      &               &             &  & MSA CA                              & \MSAcaHLsevenPlain            & \MSAcaHLsevenTLS            & \MSAcaHLsevenTotal &                 &               &             &  \\ 
                                      &             &             &  & MSA CE                             & \MSAceHLsevenPlain            & \MSAceHLsevenTLS            & \MSAceHLsevenTotal &                 &               &             &  \\ 
                                      &               &             &  & MSA CR                              & \MSAcrHLsevenPlain            & \MSAcrHLsevenTLS            & \MSAcrHLsevenTotal &                 &               &             &  \\
                                       &               &             &  & Non-compliant                           & \NoncompliantMSA            & 0            & \NoncompliantMSA &                 &               &             &  \\\bottomrule
    \end{tabular}
\end{table*}

\subsubsection{Intersection of Healthcare Services}

\begin{figure}[t]
    \centering
    \includesvg[width=\linewidth]{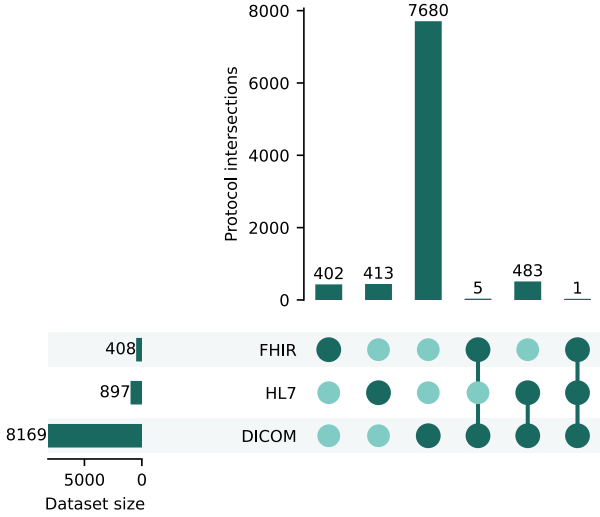}
    \caption{Number of hosts with and without intersection of multiple healthcare protocols.}
    \label{fig:multihost}
\end{figure}

As \cref{fig:multihost} shows, some systems expose more than one healthcare service to the Internet. In total, there are \MultidicomHlsevnOnly ~systems that provide both a DICOM and an HL7 service. This corresponds to \MultiHLsvenPercentage\% of all HL7 systems. This observation can be explained by the fact that DICOM applications like Picture Archiving and Communication Systems (PACS), for example, have both DICOM and HL7 interfaces. Only \MultidicomFhirOnly{} systems were found that expose both DICOM and FHIR services. Manual examination of the \MultidicomFhirOnly{} systems revealed that, unlike the combination of DICOM and HL7, this is not an application that provides both protocols, but rather several applications running simultaneously on a server. In these cases, a conceivable scenario is that the DICOM data (e.g., patient master data) is processed directly via an interface in the FHIR application. Only one host serves all three healthcare protocols. This host runs a test framework for validating IHE-compliant interfaces for healthcare protocols, which supports all three protocols.

\section{Security \& Privacy Evaluation}

\subsection{Data Leaks}

\subsubsection{DICOM Data Leaks}
For each confirmed DICOM service, we tried to negotiate an association using the default values for the calling AE-title (\textit{PYNETDICOM}) and the called AE-title (\textit{ANY-SCP}), and we checked support for DICOM C-FIND requests. In total, this led to \AllDICOMCFindSupport{} successfully established associations.
When querying patient data, our measurements show that in \AllDICOMCFindFraction\% of these cases, a successful association is sufficient to access patient data without further access control, resulting in \AllDICOMCFindData{} DICOM services leaking patient data to arbitrary clients.

In case the association failed, in \CFindFailedCalledAETitle ~out of \CFindFailedAETitleTotal ~times, the failure resulted from a rejection of the called AE-title. This means the server will not accept connections from arbitrary clients; instead, it requires them to know the correct server AE-title to connect, which effectively serves as a security/authentication feature.
This behavior motivates evaluating leaked AE-titles via LDAP servers.
Out of all confirmed DICOM services, \DICOMLDAP ~run in parallel with an LDAP server. Out of these, \DICOMLDAPtitleleak ~LDAP servers allowed us to query AE-titles without requiring authentication, resulting in a leak of the necessary access data. \cref{tab:aetitles} in the appendix lists the top 20 AE-titles that we were able to retrieve via the LDAP queries.

\subsubsection{HL7 Data Leaks}

We did not find any strings in the ERR-7 field that confirm the presence of the UAC authentication feature. This means that none of the HL7 endpoints found (n=\AllHLseven) use a genuine authentication method. In \MessageAppIssueCount ~cases, we found indications in the HL7 message that the sending Application or Facility or the receiving Application or Facility is unknown or not authorized. This could suggest a simple way to increase the security of HL7 endpoints, similar to DICOM's AE-titles.

However, this does not mean that all endpoints except for these \MessageAppIssueCount ~HL7 endpoints are affected by data leaks, as most HL7 endpoints do not support data extraction. The evaluation of the MSA segments shows that in \CodeSUCCESS ~cases, the HL7 response message contains an \texttt{AA} code. In these cases, the endpoint processed the message without generating an error. In \CodeERROR ~of cases, the response contains an \texttt{AR} or \texttt{AE}.

Nevertheless, these figures do not imply that \CodeSUCCESS ~percent of all systems are affected by a data leak. \AAEmptyCount~ HL7 messages with \texttt{AA} codes are empty; it is therefore questionable whether these messages were processed correctly. On the other hand, we also found a message with an AE code and the following text: \textit{"query is too vague, query requires either phonenumber, dateofbirth, both firstname and surname, both id and assigningauthorityid parameters"}. This indicates that the service accepts HL7 queries, but the search parameters were incorrect.

For further analysis, we divided the HL7 responses into three groups based on the content of the free text fields (MSA-3 or ERR-8): Processed (n=\MessageProcessedCount), Not processed (n=\MessageNotProcessedCount), and Unknown (n=\MessageUnknownCount). There are a total of \MessageEmptyCount ~HL7 messages with empty text fields in the unknown group. In the \MessageProcessedCount ~cases in which our request was processed, we assume that patient data can be retrieved using a correct HL7 message with an existing patient. In \MessageNotProcessedCount ~cases, the message could not be processed, meaning that no patient data could be retrieved because the endpoint does not support query messages such as PDQ.

Regarding HL7, it can be said that publicly accessible HL7 endpoints are not at great risk of data leaks, since most do not support data retrieval. The different handling of HL7 messages, which is due to the considerable design freedom granted by the HL7 standard, makes it difficult to assess whether a endpoint would ultimately disclose data if real patient names or IDs were queried. For an attacker to extract data from an HL7 endpoint, the endpoint would first have to support a query function such as PDQ. The attacker would also need to know how to structure the search query correctly for the endpoint to process it. HL7 queries do not allow the use of wildcards in search queries. 

However, manipulating data on HL7 endpoints is a far greater security concern than reading publicly accessible HL7 endpoints. Since we found only 14 endpoints with security measures such as sender and receiver checks, none of which use UAC, and since HL7 is designed for pushing messages, we assume that data could be manipulated on almost all endpoints. 

Manipulating the data can have serious consequences for patients. Therefore, every publicly accessible HL7 endpoint must be considered a major security risk.

\subsubsection{FHIR Data Leaks}

Our data shows that we found \FHIRPatientHTTPCodeOK ~FHIR endpoints that responded with an HTTP code 200 when querying the patient resource. Of these \FHIRPatientHTTPCodeOK ~endpoints, \FHIRPatientEndpointsCount ~endpoints contain unprotected patient data.

The \FHIRPatientEndpointsCount ~unprotected endpoints reveal a total of \FHIRPatientTotalCount ~patients. However, this patient data might include collisions and test data that do not reflect real patients. \cref{tab:fhir_patients} shows the 20 most common patient data per endpoint.

Another \FHIRPatientHTTPCodeUnauthorized ~FHIR endpoints returned HTTP code 401. This HTTP code implies that access was denied due to missing credentials. These systems, therefore, most likely use authentication. Another \FHIRPatientHTTPCodeInternalServerError ~FHIR endpoints returned HTTP code 500, which indicates an error in the web server. \FHIRPatientHTTPCodeNotFound ~FHIR endpoints returned the HTTP code 404, which means that the resource does not exist, and in this case an OperationOutcome error is displayed by the FHIR endpoint. This is normal behavior for FHIR endpoints when a FHIR resource is not supported. Of the remaining endpoints, \FHIRPatientHTTPCodeOther ~returned other HTTP codes or timed out.

The values for the FHIR resource \texttt{/Organization} behave similarly. When requesting this resource, \FHIROrgaHTTPCodeOK ~endpoints responded with HTTP code 200, \FHIROrgaHTTPCodeUnauthorized ~endpoints responded with HTTP code 401, \FHIROrgaHTTPCodeInternalServerError ~returned code 500, and \FHIROrgaHTTPCodeNotFound ~returned code 404. \FHIROrgaHTTPCodeOther ~returned other codes, and the remaining endpoints timed out.

In addition to the HTTP codes, we also evaluated the authentication methods offered by the FHIR endpoint via the \texttt{/Metadata} endpoint. \Cref{tab:fhir_auth} shows the number of endpoints that advertise specific combinations of three different authentication methods referenced by the queried endpoints. The given information may be flawed, incomplete, or unnecessary for resource access, as endpoints may claim authentication support even if they do not require authentication. Non-conformant metadata may be classified as not advertising any authentication. To provide further insight, we show the number of endpoints that do not require authentication to access patient data. However, an endpoint without authentication might not enable access to the \texttt{/Patient} resource if the functionality is not implemented.

\begin{table}
\caption{Top 20 FHIR Endpoints exposing patient data.}
\label{tab:fhir_patients}
\centering
\footnotesize
\begin{tabular}{lrr}
\toprule
\multicolumn{1}{c}{\textbf{Endpoints}} & \multicolumn{1}{c}{\textbf{Patients}} \\ \midrule
https://x.x.x.x:443/fhir & 685,100\\
http://x.x.x.x:9080/fhir & 585,569\\
https://x.x.x.x:443/baseR4 & 435,745\\
http://x.x.x.x:8080/fhir & 176,306\\
http://x.x.x.x:8080/fhir & 59,804\\
http://x.x.x.x:9080 & 46,658\\
http://x.x.x.x:80/fhir & 30,730\\
http://x.x.x.x:8082/fhir & 18,448\\
https://x.x.x.x:443/baseR5 & 17,635\\
http://x.x.x.x:8090/fhir & 12,915\\
http://x.x.x.x:8080/fhir & 10,567\\
http://x.x.x.x:8080/fhir & 9,498\\
http://x.x.x.x:80/fhir & 8,567\\
http://x.x.x.x:8081/fhir & 5,868\\
http://x.x.x.x:8080/fhir & 2,915\\
https://x.x.x.x:443/baseR5 & 2,122\\
https://x.x.x.x:443/baseR4 & 2,121\\
https://x.x.x.x:443/baseDstu3 & 2,115\\
http://x.x.x.x:80 & 1,482\\
http://x.x.x.x:8080/fhir & 1,275\\
\bottomrule
\end{tabular}
\end{table}

\begin{table}
    \caption{Authentication methods advertised by FHIR endpoints, categorized by access to patient data.}
    \label{tab:fhir_auth}
    \centering
    \footnotesize
    \begin{tabular}{lrr}
        \toprule
        \textbf{Method Combination}      & \textbf{No Patient Access}& \textbf{Patient Access}\\ \midrule
        none                             & 295 (58.3\%)                 & 149 (29.45\%)             \\
        smart-on-fhir                    & 24 (4.74\%)                  & 1 (0.2\%)                 \\
        basic, smart-on-fhir, oauth      & 13 (2.57\%)                  & 1 (0.2\%)                 \\
        oauth                            & 12 (2.37\%)                  & -                         \\
        basic                            & 8 (1.58\%)                   & 1 (0.2\%)                 \\
        smart-on-fhir, oauth             & 2 (0.4\%)                    & -                         \\
        \bottomrule
    \end{tabular}
\end{table}

\subsection{TLS Analysis}

The evaluation of TLS handshakes (see \cref{tab:tls}) shows that, while TLS support for healthcare protocols is generally low, it varies by protocol. Only \DICOMTLSPercentage\% of DICOM hosts support TLS at all. Of those, \DICOMTLSOneZeroPercent\% negotiated the outdated version 1.0, while a further \DICOMTLSOneOnePercent\% negotiated the equally outdated version 1.1. Overall, \DICOMInsecureCiphersPercentage\% of DICOM hosts negotiated a weak or insecure cipher suite, as indicated by~\cite{ciphersuite}. In \DICOMSelfSignedPercentage\% of cases, DICOM hosts presented a self-signed certificate. 

TLS support for HL7 is almost non-existent. Only two systems (\HLSevenTLSPercentage\%) answer to TLS connection attempts at all. These two systems expose a TLS server that is securely configured. Compared to other healthcare protocols, TLS usage is significantly higher for FHIR: In fact, \FHIRTLSPercentage\% of hosts use TLS, all of which support version 1.2, which is still considered secure. Only \FHIRInsecureCiphersPercentage\% of hosts use a weak or insecure cipher suite regarding \cite{ciphersuite}. Self-signed certificates are used by only \FHIRSelfSignedPercentage\% of hosts. None of the TLS connections tested for any of the three protocols were implemented via TLS 1.3.

\begin{table*}
    \caption{Analysis of TLS usage, version, cipher suites and self-signed certificates for healthcare protocol sockets (IP and port combination).}
    \label{tab:tls}
    \footnotesize
    \centering
    \begin{tabular}{@{}rrrrrrrrr@{}}
    \toprule
    \multicolumn{1}{c}{\textbf{Protocol}} & \multicolumn{1}{c}{\textbf{TLS Usage}} & \multicolumn{4}{c}{\textbf{TLS Version}} & \multicolumn{2}{c}{\textbf{Cipher Suites}} & \multicolumn{1}{c}{\textbf{Self Signed}} \\
                                          &  & \multicolumn{1}{c}{1.3} & \multicolumn{1}{c}{1.2} & \multicolumn{1}{c}{1.1} & \multicolumn{1}{c}{1.0} & Secure  & Insecure & \multicolumn{1}{c}{\textbf{Certificates}} \\ \midrule
    \textbf{DICOM} & \DICOMTotalHandshakes (\DICOMTLSPercentage\%)  & \DICOMTLSOneThree (\DICOMTLSOneThreePercent\%) & \DICOMTLSOneTwo (\DICOMTLSOneTwoPercent\%)  & \DICOMTLSOneOne (\DICOMTLSOneOnePercent\%) & \DICOMTLSOneZero (\DICOMTLSOneZeroPercent\%) & \DICOMSecureCiphers (\DICOMSecureCiphersPercentage\%) & \DICOMInsecureCiphers (\DICOMInsecureCiphersPercentage\%) & \DICOMSelfSignedCount (\DICOMSelfSignedPercentage\%) \\
    \textbf{HL7}  & \HLSevenTotalHandshakes (\HLSevenTLSPercentage\%)  & \HLSevenTLSOneThree (\HLSevenTLSOneThreePercent\%) & \HLSevenTLSOneTwo (\HLSevenTLSOneTwoPercent\%)  & \HLSevenTLSOneOne (\HLSevenTLSOneOnePercent\%) & \HLSevenTLSOneZero (\HLSevenTLSOneZeroPercent\%) & \HLSevenSecureCiphers (\HLSevenSecureCiphersPercentage\%) & \HLSevenInsecureCiphers (\HLSevenInsecureCiphersPercentage\%) & \HLSevenSelfSignedCount (\HLSevenSelfSignedPercentage\%) \\
    \textbf{FHIR} & \FHIRTotalHandshakes (\FHIRTLSPercentage\%)  & \FHIRTLSOneThree (\FHIRTLSOneThreePercent\%) & \FHIRTLSOneTwo (\FHIRTLSOneTwoPercent\%)  & \FHIRTLSOneOne (\FHIRTLSOneOnePercent\%) & \FHIRTLSOneZero (\FHIRTLSOneZeroPercent\%) & \FHIRSecureCiphers (\FHIRSecureCiphersPercentage\%) & \FHIRInsecureCiphers (\FHIRInsecureCiphersPercentage\%) & \FHIRSelfSignedCount (\FHIRSelfSignedPercentage\%) \\  \bottomrule
    \end{tabular}
\end{table*}

\subsection{Vendors, Versions \& CVEs}

We categorize the identified healthcare systems by vendor and version. Since healthcare protocols vary in functionality, there are differences in the depth of the evaluation.

\subsubsection{DICOM Implementations}

The analysis of the DICOM software libraries used (see \cref{tab:top10vendors}) shows that the \texttt{OFFIS DCMTK} library, in various versions, is used most frequently, in total by \DICOMDCMTKCount~DICOM endpoints. \DICOMJDTCount ~endpoints indicated the software library \texttt{jdt270\_6009}, which refers to the Java DICOM Toolkit in version 2.70. The third most frequently used implementation is the DICOM Image Library from Osirix with \DICOMOsirixCount ~endpoints.

\subsubsection{HL7 Versions}

 Of the \AllHLseven ~HL7 endpoints in our data, a total of \HLsevenVersion ~used version 2.5.1, which we also used for our PDQ message. The remaining \HLsevenVOther ~endpoints use versions between 2.2 and 2.8. An HL7 endpoint usually supports multiple versions. When a request is made using a specific version, the response is generally sent in the same version. To verify this behavior, we sent an additional message with version 2.4 to the systems. In over 94\% of cases, the HL7 server returned the same version. Endpoints that do not support HL7 version 2.6 or higher do not support UAC authentication.
    
\subsubsection{FHIR Vendors and Versions}

FHIR offers the most detailed option for determining the software vendor and version. The data shows that the open-source HAPI FHIR Server, in various versions, is the most widely used FHIR software (\FHIRAllHAPIServer). Our analysis also shows that \FHIRVersionFourOOne ~FHIR endpoints use version 4.0.1 of the FHIR standard. By contrast, \FHIRVersionFive ~endpoints use the latest version, 5.0.0, of the FHIR standard.

\renewcommand{\arraystretch}{1.2}
\begin{table}
    \caption{Top 10 used software products (DICOM and FHIR) and including security vulnerabilities.}
    \label{tab:top10vendors}
    \footnotesize
    \centering
    \begin{adjustbox}{max width=\linewidth}
    \begin{tabular}{p{3.8cm}rrr}
        \toprule
        \multicolumn{4}{c}{\textbf{DICOM}}                       \\ \midrule
        \textbf{Software and Version}     & \textbf{Count}     & \textbf{CVE Count} & \textbf{Max CVSS} \\ \hline
        jdt270\_6009 & 454 (23.7\%) & 0 & N/A \\
        OFFIS\_DCMTK\_369 & 267 (13.9\%) & 2 & 7.8 \\
        OFFIS\_DCMTK\_367 & 235 (12.3\%) & 5 & 7.8 \\
        OSIRIX\_361 & 208 (10.9\%) & 1 & 1.9 \\
        OFFIS\_DCMTK\_368 & 161 (8.4\%) & 7 & 8.1 \\
        OFFIS\_DCMTK\_366 & 139 (7.3\%) & 11 & 9.8 \\
        OSIRIX & 129 (6.7\%) & N/A & N/A \\
        OFFIS\_DCMTK\_365 & 88 (4.6\%) & 12 & 9.8 \\
        OFFIS\_DCMTK\_364 & 71 (3.7\%) & 12 & 9.8 \\
        OFFIS\_DCMTK\_361 & 61 (3.2\%) & 1 & 7.2 \\
        \toprule
        \multicolumn{4}{c}{\textbf{FHIR}}                        \\ \midrule
        \textbf{Software and Version}     & \textbf{Count}     & \textbf{CVE Count} & \textbf{Max CVSS} \\ \midrule
        Snowstorm X FHIR Server N/A & 43 (13.4\%) & N/A & N/A \\
        HAPI FHIR Server 8.4.0 & 28 (8.7\%) & 0 & N/A \\
        HAPI FHIR Server 8.2.0 & 23 (7.2\%) & 0 & N/A \\
        HAPI FHIR Server 8.0.0 & 16 (5.0\%) & 0 & N/A \\
        HAPI FHIR Server 7.4.0 & 15 (4.7\%) & 0 & N/A \\
        HAPI FHIR Server 7.0.2 & 14 (4.4\%) & 0 & N/A \\
        HAPI FHIR Server 7.6.0 & 10 (3.1\%) & 0 & N/A \\
        HAPI FHIR Server 6.10.0 & 9 (2.8\%) & 0 & N/A \\
        HAPI FHIR Server 6.2.2 & 8 (2.5\%) & 1 & 9.8 \\
        HAPI FHIR Server 5.1.0 & 8 (2.5\%) & 2 & 9.8 \\
        \bottomrule
    \end{tabular}
    \end{adjustbox}
\end{table}

\subsubsection{CVE Analysis}

Several known security vulnerabilities can be identified in the found DICOM and FHIR systems. In total, we discovered \AllServerCVE ~systems with at least one security vulnerability. Of these, \DICOMServerCVE ~DICOM and \FHIRServerCVE ~FHIR systems are affected. \CVETwentyTwoTwoOneOneNine ~DICOM systems use the OFFIS DCMTK Library in a version before 3.6.7 and are therefore affected by the two security vulnerabilities CVE-2022-2119 and CVE-2022-2120, each with a CVSS score of 9.8. Both vulnerabilities allow an attacker to write DICOM files and perform remote code execution. \CVETwentyTwentyFourFiveOneOneThree ~FHIR systems are affected by the CVE-2024-51132 vulnerability, which also has a CVSS score of 9.8. This vulnerability can be exploited to retrieve sensitive information or execute malicious code.

\subsubsection{Global Exposure}

\renewcommand{\arraystretch}{1.2}
\begin{table*}
\caption{The top 20 countries worldwide with the highest number of exposed DICOM, HL7, and FHIR unique hosts, including those affected by data leaks, TLS flaws, and known vulnerabilities.}
\label{tab:geo_hl7_dicom_fhir}
\centering
\begin{adjustbox}{max width=\linewidth}
\begin{tabular}{lrrrrrrrrrrrrr}
\toprule
\multirow{2}{*}{\textbf{Region}} &
\multirow{2}{*}{\textbf{Hosts}} &
\multirow{2}{1.5cm}{\textbf{Hosts per 1M Inhab.}} &
\multicolumn{4}{c}{\textbf{DICOM}} &
\multicolumn{3}{c}{\textbf{HL7}} &
\multicolumn{4}{c}{\textbf{FHIR}} \\ 
\cmidrule(lr){4-7} \cmidrule(lr){8-10} \cmidrule(lr){11-14}
 &  & &
\multicolumn{1}{c}{\textbf{Hosts}} &
\multicolumn{1}{c}{\textbf{Leaks}} &
\multicolumn{1}{c}{\textbf{CVE}} &
\multicolumn{1}{c}{\textbf{TLS Flaws}} &
\multicolumn{1}{c}{\textbf{Hosts}} &
\multicolumn{1}{c}{\textbf{Leaks}} &
\multicolumn{1}{c}{\textbf{TLS Flaws}} &
\multicolumn{1}{c}{\textbf{Hosts}} &
\multicolumn{1}{c}{\textbf{Leaks}} &
\multicolumn{1}{c}{\textbf{CVE}} &
\multicolumn{1}{c}{\textbf{TLS Flaws}} \\ 
\midrule
US & 2377 (26.4\%)& 6,99 & 2107 (23.4\%) & 262 (2.9\%) & 261 (2.9\%) & 87 (1.0\%) & 221 (2.5\%) & 5 (0.1\%) &  0 & 158 (1.8\%) & 43 (0.5\%) & 22 (0.2\%) & 6 (0.1\%) \\
BR & 1025 (11.4\%)& 4,83 & 981 (10.9\%) & 28 (0.3\%) & 110 (1.2\%) & 0 & 107 (1.2\%) & 11 (0.1\%) &  0 & 4 (0.0\%) & 1 (0.0\%) & 0 & 0 \\
IN & 819 (9.1\%)& 0,56 & 764 (8.5\%) & 35 (0.4\%) & 217 (2.4\%) & 0 & 98 (1.1\%) & 1 (0.0\%) &  0 & 24 (0.3\%) & 10 (0.1\%) & 2 (0.0\%) & 0 \\
CN & 310 (3.4\%)& 0,22 & 297 (3.3\%) & 30 (0.3\%) & 88 (1.0\%) & 0 & 9 (0.1\%) & 0 &  0 & 9 (0.1\%) & 3 (0.0\%) & 0 & 0 \\
VN & 266 (3.0\%)& 2,63 & 256 (2.8\%) & 2 (0.0\%) & 29 (0.3\%) & 0 & 51 (0.6\%) & 0 &  0 & 2 (0.0\%) & 0 & 0 & 0 \\
FR & 262 (2.9\%)& 3,91 & 225 (2.5\%) & 69 (0.8\%) & 61 (0.7\%) & 0 & 33 (0.4\%) & 0 &  0 & 31 (0.3\%) & 8 (0.1\%) & 0 & 0 \\
CA & 219 (2.4\%)& 5,34 & 157 (1.7\%) & 35 (0.4\%) & 17 (0.2\%) & 1 (0.0\%) & 54 (0.6\%) & 2 (0.0\%) &  0 & 13 (0.1\%) & 2 (0.0\%) & 1 (0.0\%) & 1 (0.0\%) \\
IR & 189 (2.1\%)& 2,05 & 189 (2.1\%) & 2 (0.0\%) & 18 (0.2\%) & 22 (0.2\%) & 0 & 0 &  0 & 0 & 0 & 0 & 0 \\
EG & 189 (2.1\%)& 1,62 & 180 (2.0\%) & 0 & 8 (0.1\%) & 0 & 16 (0.2\%) & 0 &  0 & 0 & 0 & 0 & 0 \\
DE & 183 (2.0\%)& 2,18 & 156 (1.7\%) & 75 (0.8\%) & 32 (0.4\%) & 2 (0.0\%) & 9 (0.1\%) & 0 &  0 & 23 (0.3\%) & 10 (0.1\%) & 6 (0.1\%) & 0 \\
GB & 161 (1.8\%)& 2,33 & 133 (1.5\%) & 58 (0.6\%) & 37 (0.4\%) & 1 (0.0\%) & 10 (0.1\%) & 1 (0.0\%) &  0 & 20 (0.2\%) & 4 (0.0\%) & 0 & 0 \\
CL & 140 (1.6\%)& 7,00 & 113 (1.3\%) & 11 (0.1\%) & 13 (0.1\%) & 0 & 37 (0.4\%) & 0 &  0 & 3 (0.0\%) & 1 (0.0\%) & 0 & 0 \\
ZA & 131 (1.5\%)& 2,05 & 123 (1.4\%) & 12 (0.1\%) & 6 (0.1\%) & 9 (0.1\%) & 5 (0.1\%) & 1 (0.0\%) &  0 & 3 (0.0\%) & 1 (0.0\%) & 0 & 0 \\
AR & 129 (1.4\%)& 2,80 & 110 (1.2\%) & 4 (0.0\%) & 17 (0.2\%) & 0 & 33 (0.4\%) & 3 (0.0\%) &  0 & 2 (0.0\%) & 1 (0.0\%) & 0 & 0 \\
RU & 123 (1.4\%)& 0,85 & 121 (1.3\%) & 13 (0.1\%) & 30 (0.3\%) & 1 (0.0\%) & 8 (0.1\%) & 0 &  0 & 4 (0.0\%) & 0 & 0 & 0 \\
CO & 122 (1.4\%)& 2,30 & 120 (1.3\%) & 4 (0.0\%) & 20 (0.2\%) & 0 & 16 (0.2\%) & 0 &  0 & 0 & 0 & 0 & 0 \\
MX & 118 (1.3\%)& 0,90 & 103 (1.1\%) & 5 (0.1\%) & 16 (0.2\%) & 0 & 26 (0.3\%) & 0 &  0 & 0 & 0 & 0 & 0 \\
ID & 115 (1.3\%)& 0,40 & 111 (1.2\%) & 37 (0.4\%) & 33 (0.4\%) & 0 & 9 (0.1\%) & 0 &  0 & 1 (0.0\%) & 0 & 0 & 0 \\
KR & 112 (1.2\%)& 2,15 & 108 (1.2\%) & 12 (0.1\%) & 22 (0.2\%) & 0 & 1 (0.0\%) & 0 &  0 & 3 (0.0\%) & 2 (0.0\%) & 0 & 0 \\
NL & 102 (1.1\%)& 5,67 & 84 (0.9\%) & 33 (0.4\%) & 10 (0.1\%) & 0 & 7 (0.1\%) & 0 &  0 & 14 (0.2\%) & 4 (0.0\%) & 0 & 0 \\
\midrule
Total: & 8987 (100.0\%)& Sum: & 8169 (90.9\%) & 1129 (12.6\%) & 1320 (14.7\%) & 141 (1.6\%) & 897 (10.0\%) & 27 (0.3\%) &  0 & 408 (4.5\%) & 130 (1.4\%) & 40 (0.4\%) & 8 (0.1\%) \\

\bottomrule

\end{tabular}
\end{adjustbox}
\end{table*}

\Cref{tab:geo_hl7_dicom_fhir} compares the 20 countries with the most exposed DICOM, HL7, and FHIR hosts. The data is based on unique IP addresses rather than endpoints, as endpoints with the same IP address would otherwise be assigned to the same country multiple times. The analysis reveals clear regional disparities in healthcare protocol security.

The United States has by far the most significant exposure, accounting for over a quarter (26.4\%) of all identified hosts, and it shows numerous vulnerabilities across all three protocols, especially in DICOM. Brazil and India follow, each representing approximately 10\% of total hosts. Notably, India has a higher number of CVEs per host than Brazil. In contrast, European countries such as France and Germany exhibit a lower total number of exposures but a comparatively higher proportion of data leaks in FHIR and DICOM systems.

\section{Responsible Disclosure}

To mitigate the identified security risks, we first reached out to a national CERT, with the goal of leveraging an established Coordinated Vulnerability Disclosure procedure. 
In response, the CERT informed us that it does not consider itself responsible for the disclosure procedure. 
We therefore carried out the responsible disclosure ourselves: 
We determined the abuse information for each vulnerable endpoint via Censys.io and reached out to the responsible operators. We instructed large service providers such as hosting providers to forward the information the relevant administrators/organizations.
We also attached a short questionnaire to our disclosure emails (see Appendix \ref{sec:apendix_Disclosure}). 
A total of 5,806 hosts were included in the disclosure and sent to 550 different email addresses.

After a 4-week waiting period, we repeated the scans for all three healthcare protocols.
For DICOM, we observed that \DICOMDisclosureRemoved~DICOM hosts were not publicly accessible anymore. However, we found \DICOMDisclosureNew~new DICOM hosts, indicating that a large part of identified hosts may change their IP addresses dynamically, e.g., due to ISP settings. Ultimately, we measured a total difference of \DICOMDisclosureDiff~hosts, which equals to a change of \DICOMDisclosurePercent\%.

Similarly, 
HL7 scans revealed that \HLSevenDisclosureRemoved ~hosts were no longer present, while \HLSevenDisclosureNew ~new hosts were identified. This represents a reduction of \HLSevenDisclosurePercent\%. With regard to FHIR, 228 hosts can no longer be found, while 163 new ones have been added. This represents a reduction of -15.08\%.

A total of just 26 people responded to the questionnaire. 25 respondents answered 'yes' to Question 1, and 24 also answered 'yes' to Question 2. Therefore, it can be concluded that the systems are intentionally made available on the internet. However, only 4 of 22 respondents stated that they reconfigured the systems accordingly (Question 3). When asked which contact method was best (Question 4), most respondents said that an abuse email address or alternative email address was best. 13 people voluntarily indicated in the free-text field for Question 4 that the system in question was a honeypot.

Overall, we found the disclosure procedure had little impact on reducing the total amount of publicly available, vulnerable systems. The number of systems before and after the procedure changed only minimally, and the low response rate to our questionnaire is further evidence of this.

\section{Discussion}
\label{sec:discussion}

Our results show that a large share of publicly exposed healthcare services (via DICOM, HL7, or FHIR) suffers from security flaws.
DICOM and HL7 endpoints that expose patient data are usually the result of incorrect network segmentation, as these endpoints should not be reachable via the Internet.
Furthermore, our analyses show that encryption is rarely available for the transmission of medical data, and various endpoints suffer from critical security vulnerabilities and missing authentication.

Yet the consequences of such security gaps are not merely technical. Patient safety is the most valuable asset of healthcare facilities, but it may be compromised by inadequate security measures. In some of the cases presented, attackers may be able to manipulate patient data, posing a serious risk to patients.

Some of the security deficiencies we identified may be explained by the special demands on healthcare systems. For example, the complexity of establishing and operating transport encryption may pose a risk to patient treatment. Expired TLS certificates can cause connection establishment to fail, leading to communication loss between medical devices and unpredictable consequences. 
Against this background, the benefits of deploying \textit{in-house} transport encryption may not outweigh their risks.

However, these \textit{internal} systems should never be accessible via the Internet, and it seems operators might be unaware of this. %
Our results suggest that operators are having difficulty with network configuration and do not check their own infrastructure for exposure; otherwise, these systems would be identified and the issues resolved. 
Similarly, it is questionable whether healthcare institutions are aware that they are using software with known vulnerabilities or the risks they are exposing themselves to.
Reasons for that may be manifold; however, it is well known that cybersecurity in healthcare has been chronically underfunded, leaving the sector particularly exposed~\cite{Cartwright2023}. Additionally, healthcare environments use complex, non-mainstream technologies such as HL7 and DICOM, which may make their security more challenging~\cite{jalali2018}. 

As a first step towards securing their environments, healthcare facilities should introduce a strict policy that medical systems should never be connected to the internet, and implement the highest possible regulatory and technical barriers to prevent this.
We furthermore propose that healthcare institutions reduce the attack surface as much as possible, particularly by implementing recurring vulnerability scans as a cost-effective, scalable mitigation. 
In practice, operators can use network scans, security scanners, and audits of authentication methods to detect and remedy all of the security issues we have described. To facilitate such scans, we are making the scanners presented in this paper available as open source.

FHIR is the only healthcare protocol designed to be exposed to the Internet for data exchange between parties. As FHIR is based on modern web technology, well-known security methodologies can be applied. Moreover, FHIR itself defines a special authentication method (SMART-on-FHIR) based on OAuth2.
DICOM can, in theory, be deployed securely on the Internet, as it includes relevant security features in Parts 7 and 15 of the standard. However, in practice, these security features are not deployed by healthcare facilities and often not even implemented in software. The most often deployed \enquote{security measure} for DICOM, the verification of AE-titles, is not a sufficient authentication method. AE-titles are strings of up to 16 characters intended for identifying medical software or devices. Often, AE-titles can be obtained via LDAP or via the software documentation, which is usually available online. This makes it very easy to brute force or infer AE-titles.
HL7 defines no security features (except for UAC) and should therefore never be made available on the Internet.

\subsection{Limitations}
\label{sec:limits}

Our study has some limitations that affect the results. First, it is worth noting that we only scan from a single server in a single location (Germany). As Wan et al.~\cite{geoscan} demonstrated, scanning from a single location can negatively affect results, as not all systems can be reached due to network issues and geo-blocking. In accordance with ethical considerations, our Internet scans are based on an opt-out procedure. We added IP ranges of systems whose operators disagree with the scans to an internal scan block list. It is also possible for operators to block our scan server in their firewalls to prevent our scans. The use of different ports by operators can lead to further uncertainty. This applies particularly to FHIR and HL7, as multiple ports are used for these protocols. Limited time and technical resources make it impossible to scan all ports. It should also be noted that it is not possible to identify all active IPv6 addresses, as previous studies have shown~\cite{10.1145/3749215}. Even combining several methods does not guarantee that all IPv6 hosts will be found. Therefore, our scanning results provide a lower bound estimate on the number of healthcare services available online.

In addition to network-level limitations, there are also protocol-level limitations. When identifying FHIR systems, we need to know the FHIR URL. However, as already described, the FHIR URL can usually be freely chosen by the operator. Therefore, it is not possible to track down all FHIR endpoints. A FHIR system can also remain undetected if the actual public metadata (CapabilityStatement) cannot be retrieved due to an incorrect implementation of the FHIR system or the use of authentication for metadata access.

Furthermore, only the web server’s standard certificate is analysed, and only the CN is read. If the FHIR server can only be accessed via the domain name (i.e. virtual host and Server Name Indication) and does not have a TLS certificate, or if the correct system name is only listed in the Subject Alternative Name field or is not included in the standard certificate, the FHIR endpoint will remain unrecognised. 
With regard to HL7, we have refrained from searching for HL7v3 systems, as this version is not widely used and, in our opinion, the effort required to do so is disproportionate.

The limitations mentioned above may mean that some systems are not found, and that the number of healthcare systems we identified could be higher.

For ethical reasons, we do not download or review patient data that might be present on identified systems. Therefore, we cannot evaluate which percentage of the found systems are test systems, honeypots, or production systems, and whether they contain real patient data.  Even if we could identify test systems, it is unclear whether they contain real patient data. It is therefore not possible to draw a definitive conclusion about the systems found.

\section{Ethical Considerations}
\label{sec:ethics}

Handling sensitive patient data must be carried out in strict accordance with ethical principles to avoid harming patients. The same applies to systems that process this data and are directly or indirectly involved in patient treatment.

To ensure compliance with all ethical considerations, we conducted a stakeholder-based ethical analysis and applied the principles of the Menlo Report to our research~\cite{menlo2012}. At the beginning of our research, we identified the following stakeholders: Patients, Malicious Actors, Healthcare facilities, IT staff and IT managers at healthcare facilities, Society, and Network and platform operators.

Patients may be affected by the research, as their data could potentially be accessed from the Internet without authentication or even manipulated. To protect patient safety, we have taken great care throughout the research process to avoid downloading any real data or manipulating it in any way. We are convinced that our research can help reduce the number of unprotected patient data on the Internet by informing the operators of medical facilities about potential security issues through our disclosure procedure. Our research, therefore, has exclusively positive effects on patients.

Malicious actors could use our research to attack vulnerable systems. However, these attacks are not new, and the findings on vulnerable DICOM systems are well known. The risk of misuse of results for HL7 and FHIR is minimally higher, as no results from large-scale scans have yet been published for these protocols. However, HL7, like DICOM, has been an established protocol for a long time, and its lack of security features is well known. It can therefore be assumed that potential attackers are also aware of this. FHIR is intended for use on the Internet, and the authentication methods of some FHIR servers can be easily supplemented. We therefore believe that our research will raise awareness of security issues among FHIR system operators, thereby improving patient safety. Our findings will only be published after the disclosure process, which further minimizes the likelihood of attacks on the systems. Furthermore, we do not publish any information that could be used to identify the vulnerable systems. Overall, we believe that our research reduces rather than increases the risk of attacks on the systems.

Healthcare facilities are also affected by our findings, as they are responsible for the security of patient data. The companies responsible for the data leaks can be identified via IP addresses, abuse contacts, or other metadata. If this information is published, it could damage the company's image and cause patients to lose trust in the company. We have therefore decided not to publish any information that could be used to identify the Healthcare facilities. Only data necessary for the disclosure process is collected. The data is processed securely and deleted as soon as the disclosure process is complete. We therefore believe that our research does not have negative consequences for healthcare facilities.

IT staff and IT managers at healthcare facilities are listed as separate stakeholders. According to our findings, some healthcare systems have configuration flaws, and some lack security updates. It is the responsibility of IT staff and IT managers to ensure that these problems do not arise in the first place. Presenting our findings to companies could reveal internal quality issues within their IT departments. This, in turn, could have negative consequences for employees. However, the misconfigurations constitute severe deficits that have a direct negative impact on patient safety. Patient safety is the highest priority objective of a healthcare facility and must therefore be guaranteed with the highest priority. After weighing the interests of patient safety against the negative consequences for individual employees, we believe patient safety outweighs these considerations.

Society is also affected by our research. The publication could lead people to adopt a negative view of healthcare facilities and to lose trust in them, resulting in fewer visits. At the same time, society benefits from our research, as we help make systems safer, protecting potential patients. Compared to all healthcare facilities worldwide, only a very small number have been affected by data leaks. We therefore believe that society as a whole will not lose confidence in healthcare facilities and that the benefits outweigh the disadvantages.

Our global Internet scans also affect network and platform operators who run systems on the Internet. Port scans and other requests originating from our system during our research may affect other systems. To minimize risk, we have implemented all eight best-practice procedures outlined by Hantke et al. for our Internet scans~\cite{10646650}. 

With regard to the ethical principles of fairness and equity outlined in the Menlo Report, we can state that we have established a global Internet scan that searches exclusively for technical medical protocols. All Internet users are treated equally. The same applies to the largely fully automated evaluation of the results. No stakeholders are given preferred or disadvantaged treatment.

The fourth principle, Respect for Law and Public Interest, is also taken into account. Since we do not download or modify patient data, we act in accordance with the law, and our scans are designed to be transparent, as per Hantke et al. \cite{10646650} The host that performs the scans discloses all information via a website. 

The procedure described was approved by our IRB before the start of the research.

\section{Related work}
\label{sec:related_work}

Related work on healthcare protocols can be divided into two categories: scanning and security vulnerabilities.

\subsection{Scanning}

Previous studies on publicly exposed health protocols on the Internet have focused exclusively on DICOM. In 2016, Stites et al.~\cite{doi:10.2214/AJR.15.15283} conducted an extensive, Internet-wide scan to identify accessible DICOM Systems. They developed a tool called DPing and found 2,774 DICOM systems on the Internet, of which 719 were completely unprotected and could have been used in a DICOM exchange.

In 2018, Beek~\cite{McAfeeDicom} searched for DICOM data using public search engines such as shodan.io. The investigation identified numerous DICOM systems that had significant vulnerabilities. %

In 2019, security researchers from Greenbone~\cite{Greenbone1} conducted a further security analysis of publicly available DICOM systems on the Internet. They also used public sources such as Shodan.io and Censys.io and identified 2,300 DICOM systems. Of these, 1,700 DICOM systems were protected, meaning that no DICOM data could be downloaded from these systems. The DICOM systems were also enriched with additional information to enable assignment to specific countries. In total, the security researchers were able to retrieve 24.3 million DICOM studies and an estimated 399.5 million individual DICOM images. The security researchers~\cite{Greenbone2} repeated the analysis after 60 days and came to a mixed conclusion. The total number of DICOM studies found rose to 35 million, while in 11 countries, all open DICOM systems from the Internet were removed. 

In 2020, Vidal et al.~\cite{chile} also used shodan.io to search for open DICOM systems, especially in Chile. They identified 47 \gls{PACS} with 23,775 registered patients on these systems.

Another important contribution was made by Yazdanmehr et al.~\cite{Blackhat2023}, who also conducted an Internet-wide scan in 2023 to identify DICOM systems. Their results showed that 3,806 DICOM systems exposed over 59 million patient records on the Internet.

In contrast, Klick et al.~\cite{german_hospital} took a different approach. Rather than searching for specific healthcare protocols, they examined the attack surface of German hospitals. Their methodology involved first identifying the hospitals and then searching for specific services. A total of 89 distinct ports were identified, each running various services, some of which exhibited vulnerabilities. However, the results did not indicate any DICOM, HL7, or FHIR services.

\subsection{Security Vulnerabilities}

In 2018, Dameff et al. demonstrated how a MitM attack on HL7 could be carried out~\cite{Blackhat2018}. They developed a tool to eavesdrop on HL7 messages. The lack of security features in HL7 enabled the researchers to read and even manipulate HL7 messages. 

In 2019, Ortiz demonstrated a security vulnerability in the design of DICOM files~\cite{PEDICOM}. It is possible to combine a DICOM file with a PE file. The DICOM file remains completely intact and can be read by appropriate programs. At the same time, the newly created PEDICOM file can be executed directly. By inserting PE files, malware can easily be injected into a DICOM file. 

Also in 2019, Mirsky et al. successfully modified the content of a DICOM image~\cite{236284}. They were able to modify DICOM files in such a way that indications of lung cancer could be removed or injected. The researchers conclude that DICOM security features, such as encryption and digital signatures, are the most effective protective measures against this type of attack.

The above-mentioned findings by Yazdanmehr et al.~\cite{Blackhat2023} not only revealed publicly accessible DICOM systems, but also exposed an SQL injection vulnerability. To do this, the SQL payload was integrated into a C-FIND command and executed on the remote system.

\section{Conclusion}
\label{sec:conclusion}

Using a honeypot, we demonstrated that only publicly accessible DICOM systems have been actively searched for on the Internet until now. We then conducted a study of the three most essential healthcare protocols — DICOM, HL7, and FHIR — searching the Internet for accessible endpoints. Our measurements of healthcare protocols on the publicly reachable internet are the first to consider the IPv6 address space. In this paper, we present the results of the first study on public HL7 and FHIR systems. Our data shows that there are \AllHLseven ~HL7 endpoints on the Internet, \AllHLsevenwithoutauth ~of which do not use any form of authentication. We also found \FHIRPatientEndpointsCount ~FHIR endpoints that do not use authentication and leak data from a total of \FHIRPatientTotalCount ~patients. Compared to existing research, we can expand on the results for DICOM by finding a total of \AllDICOMCFindData ~systems that are leaking data and demonstrating that even systems with AE-titles are less secure than previously assumed.

Additionally, we demonstrate that many healthcare systems either do not use TLS or use insecure configurations of it, and that several systems are affected by critical software vulnerabilities. Overall, our data reveal an alarming state of cybersecurity in healthcare deployments, particularly given the possibility of patient data manipulation. We believe that this study has made an essential contribution to healthcare system security and that research into this area should be intensified.

\section*{Acknowledgments}
We would like to thank Alexander Miethe for his support in setting up the IPv6 scanning pipeline. We would also like to thank Censys for providing us with a Researcher account. We thank the anonymous reviewers for their great feedback on the paper and our shepherd for supporting us in preparing the current version.

This research work was supported by the National Research
Center for Applied Cybersecurity ATHENE.
This research was also supported by the Graduate School for Applied Research in North Rhine-Westphalia (Graduate School NRW).

The FHIR ZGrab module and several Python scripts for the scan pipeline and data analysis were developed in collaboration with the AI models Claude 4.5 and GPT 5.

\let\oldbibliography\thebibliography
\renewcommand{\thebibliography}[1]{%
  \oldbibliography{#1}%
  \setlength{\itemsep}{1.5ex}
}

\bibliographystyle{IEEEtran}
\bibliography{references}

\clearpage
\appendix
\section{Appendix}

\subsection{Data Availability \& Open Science}

The source code for the honeypot, Zgrab (including the modules for DICOM, HL7 and FHIR), the pipeline running scripts, and the TLS and CVE analysis scripts are published on GitHub (\url{https://github.com/FHMS-ITS/Healthcare-Scanning}) under an open-source licence. 

For ethical reasons, we will not publish any data we have generated on healthcare endpoints, as this information contains details of vulnerable systems that could be exploited by malicious actors.  

\subsection{Healthcare Ports and Scan Results}

The ports used by healthcare protocols are important for the results of Internet scans. Although there are specified ports for HL7 and DICOM, some manufacturers or operators deviate from these specifications. This is often unavoidable, especially with HL7, as each HL7 listener and sender requires its own port and medical integration engines often have multiple HL7 listeners and senders. The ports are then often assigned arbitrarily by the operators. To identify the ports, we therefore used various sources, which we present in Table \ref{tab:scannedports}. The scan results with the number of responsive IP addresses per port can be seen in Table \ref{tab:scan_results_ports}.

\begin{table*}[ht]
\caption{Number of responsive IP addresses broken down by port and IP version.}
\label{tab:scan_results_ports}
\footnotesize
\centering
\begin{tabular}{crrcrrcrr}
\toprule
\multicolumn{3}{c}{\textbf{DICOM}} & \multicolumn{3}{c}{\textbf{HL7}} & \multicolumn{3}{c}{\textbf{FHIR}} \\
\midrule
\textbf{Port} & \textbf{IPv4} & \textbf{IPv6} & \textbf{Port} & \textbf{IPv4} & \textbf{IPv6} & \textbf{Port} & \textbf{IPv4} & \textbf{IPv6} \\
\midrule
104 & 2,953,413 & 7,313 & 1337 & 3,912,462 & 48,196 & 80 & 47,908,945 & 1,395,729 \\
2761 & 3,170,334 & 31,166 & 2020 & 3,306,034 & 48,807 & 443 & 43,612,554 & 1,254,725 \\
2762 & 3,228,247 & 33,173 & 2575 & 3,137,846 & 31,951 & 3000 & 4,415,112 & 63,786 \\
4242 & 3,070,276 & 9,048 & 3001 & 3,575,120 & 53,230 & 8000 & 7,386,535 & 69,548 \\
4498 & 3,247,198 & 34,073 & 6660 & 2,877,805 & 7,359 & 8080 & 9,316,742 & 141,843 \\
11112 & 3,134,792 & 14,469 & 6661 & 3,056,364 & 34,739 & 8081 & 5,523,293 & 73,937 \\
 &  &  & 6667 & 3,039,097 & 8,400 & 8082 & 4,466,124 & 57,336 \\
 &  &  & 6668 & 2,887,471 & 7,640 & 8083 & 3,848,766 & 57,677 \\
 &  &  & 8100 & 3,467,155 & 52,821 & 8085 & 9,504,375 & 57,904 \\
 &  &  & 12000 & 3,299,463 & 34,961 & 8089 & 9,895,522 & 328,999 \\
 &  &  & 22222 & 3,082,730 & 8,732 & 8090 & 3,979,744 & 50,395 \\
 &  &  &  &  &  & 8888 & 5,057,571 & 66,988 \\
 &  &  &  &  &  & 9000 & 6,625,607 & 54,971 \\
 &  &  &  &  &  & 9080 & 4,459,814 & 48,262 \\
\midrule
 & 18,804,260 & 129,242 &  & 35,641,547 & 336,836 &  & 166,000,704 & 3,722,100 \\
\midrule
\textbf{Total} & \multicolumn{2}{c}{18,933,502} & & \multicolumn{2}{c}{35,978,383} & & \multicolumn{2}{c}{169,722,804} \\
\bottomrule
\end{tabular}
\end{table*}

\begin{table}[H]
\caption{List of all identified healthcare ports, including their origin.}
\label{tab:scannedports}
\centering
\scalebox{0.85}{
\begin{tabular}{lll}
\toprule
\textbf{Port}  & \textbf{Protocol} & \textbf{Origin} \\ \midrule
104   & DICOM & IANA Registration \\
2761  & DICOM & IANA Registration \\
2762  & DICOM & IANA Registration \\
4242  & DICOM & Default Orthanc Port \cite{PortApp01} \\
4498  & DICOM & Default JIVEX Port \cite{PortApp02} \\
11112 & DICOM & IANA Registration \\
1337  & HL7   & Default Symmetry PACS Port \cite{PortApp03} \\
2020  & HL7   & Default ARDIS2 Port \cite{PortApp04} \\
2575  & HL7   & IANA Registration \\
3001  & HL7   & Default Sante Worklist Server Port \cite{PortApp05} \\
6660  & HL7   & Default Hermetech HL7 Port \cite{PortApp06} \\
6661  & HL7   & Default MirthConnect Port \cite{PortApp07} \\
6667  & HL7   & Also frequently used by MirthConnect\\
6668  & HL7   & Also frequently used by MirthConnect \\
8100  & HL7   & Default CharruaPACS Port \cite{PortApp08} \\
12000 & HL7   & Default HL7 DICOM Grid Port \cite{PortApp09}  \\
22222 & HL7   & Default HL7 Soup Port \cite{PortApp10} \\
80    & FHIR  & Default HTTP Port \\
443   & FHIR  & Default HTTPS Port \\
3000  & FHIR  & Censys \& Shodan \\
8000  & FHIR  & Censys \& Shodan \\
8080  & FHIR  & Default HAPI FHIR Port \cite{PortApp11}  \\
8081  & FHIR  & Censys \& Shodan \\
8082  & FHIR  & Censys \& Shodan \\
8083  & FHIR  & Censys \& Shodan \\
8085  & FHIR  & Censys \& Shodan \\
8089  & FHIR  & Censys \& Shodan \\
8090  & FHIR  & Censys \& Shodan \\
8888  & FHIR  & Censys \& Shodan \\
9000  & FHIR  & Censys \& Shodan \\
9080  & FHIR  & Censys \& Shodan \\ \bottomrule
\end{tabular}
}
\end{table}

\subsection{AE-Titles}

Some DICOM applications use their own internal Lightweight Directory Access Protocol (LDAP) to store configuration information. This includes the AE-titles used. The LDAP systems that we were able to query without authentication contain multiple AE-titles. The 20 most common AE titles are listed in Table \ref{tab:aetitles}. It seems that open LDAP systems without authentication also contain configuration errors similar to those found in the public DICOM endpoint. 

\begin{table}[H]
\centering
\footnotesize
\caption{Top 20 most frequently leaked DICOM AE-titles.}
\label{tab:aetitles}
\begin{tabular}{rlr}
\toprule
Rank & AE-Title & Count \\
\midrule
1 & \texttt{DCM4CHEE} & 972 \\
2 & \texttt{IOCM\_REGULAR\_USE} & 562 \\
3 & \texttt{AS\_RECEIVED} & 544 \\
4 & \texttt{IOCM\_QUALITY} & 544 \\
5 & \texttt{IOCM\_WRONG\_MWL} & 544 \\
6 & \texttt{IOCM\_PAT\_SAFETY} & 544 \\
7 & \texttt{IOCM\_EXPIRED} & 542 \\
8 & \texttt{WORKLIST} & 236 \\
9 & \texttt{STORESCP} & 45 \\
10 & \texttt{GOMEDICAL} & 40 \\
11 & \texttt{ARTEM} & 32 \\
12 & \texttt{AUTANA} & 19 \\
13 & \texttt{VIEWMED} & 18 \\
14 & \texttt{NOVAPACS} & 18 \\
15 & \texttt{GOTELEMEDICINA} & 12 \\
16 & \texttt{SCHEDULEDSTATION} & 10 \\
17 & \texttt{PACS2} & 10 \\
18 & \texttt{ARC1} & 9 \\
19 & \texttt{CM\_QUILICURA} & 9 \\
20 & \texttt{XMED} & 9 \\
\bottomrule
\end{tabular}
\end{table}

\subsection{HL7 Message}

\begin{figure}[h]
    \centering
    \includegraphics[width=1\linewidth]{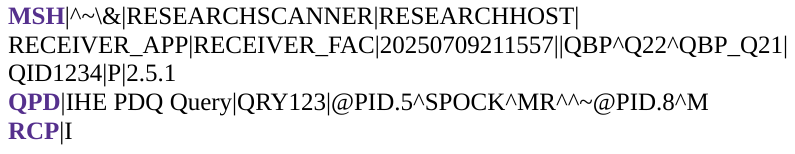}
    \caption{HL7 Message send by the HL7 Zgrab module}
    \label{fig:hl7_msg}
\end{figure}

An HL7 message usually consists of several segments. A label is used at the beginning of each segment. In Figure \ref{fig:hl7_msg}, all segment titles are highlighted in color. Typically, each HL7 message begins with the Message Header Segment (MSH). The MSH segment is used to transmit metadata for processing a message.

Depending on the type, an HL7 message can consist of several segments. Each segment in turn consists of several fields. The individual fields are separated by the \texttt{\textbar} character. The message we use is a Patient Demographics Query (PDQ), which is defined by the Query Parameter Definition (QPD) segment. In an HL7 message, the character \texttt{\textasciicircum} is used to separate multiple components within a field, for example to specify first and last names. 

\subsection{FHIR Endpoints}
\label{sec:FHIREndpoints}

We use the endpoints shown in Table \ref{tab:FhirEndpoints} to identify FHIR endpoints. Some of the endpoints contain names that depend on the FHIR version. The endpoint can therefore be used to directly identify the FHIR version used. The FHIR standard defines standard levels for the different versions. A FHIR version with the designation Normative is considered an ANSI-approved standard. The designation Trail Use is used for versions that have already been extensively tested and can also be used for productive operation. The designation Draft is used for versions that have not yet been sufficiently tested and should not be used productively.

\begin{table}[H]
\caption{The FHIR endpoints we use for identifying FHIR services, including background information on the versions used.}
\label{tab:FhirEndpoints}
\centering
\footnotesize
\begin{adjustbox}{max width=\linewidth}
\begin{tabular}{lll}
\toprule
\textbf{Endpoint} & \textbf{Version} & \textbf{Standard Level} \\ 
\midrule
\texttt{/metadata} & Version-independent &  \\
\texttt{/fhir/metadata} & Version-independent &  \\
\texttt{/baseR2/metadata} & 1.0 (R2) & Draft Standard for Trial Use 2 \\
\texttt{/baseR3/metadata} & 2.0 (R3) & Standard for Trial Use 3 \\
\texttt{/baseR4/metadata} & 4.0 (R4) & Normative\\
\texttt{/baseR5/metadata} & 5.0 (R5) & Standard for Trial Use \\
\texttt{/fhir-server/api/v4/metadata} & Version-independent &  \\
\texttt{/r5/metadata} & 5.0 (R5) & Standard for Trial Use 4 \\
\texttt{/r4/metadata} & 4.0 (R4) & Normative \\
\texttt{/r3/metadata} & 2.0 (R3) & Standard for Trial Use 3 \\
\texttt{/r2/metadata} & 1.0 (R2) & Draft Standard for Trial Use 2 \\
\texttt{/baseDstu2/metadata} & 1.0 (R2) & Draft Standard for Trial Use 2 \\
\texttt{/baseDstu3/metadata} & 2.0 (R3) & Standard for Trial Use 3 \\ 
\bottomrule
\end{tabular}
\end{adjustbox}
\end{table}

\subsection{Lantern Dataset Analysis}
\label{sec:Lanter}

Lantern \cite{Lantern1} is an open-source project designed to monitor and analyse specific FHIR endpoints. It uses public lists compiled by US health IT system developers as its data source. These lists are primarily sourced from the Certified Health IT Product List (CHPL) database \cite{Lantern2}. Lantern provides an API that allows users to download all its listed endpoints. Our analysis and comparison of Lantern’s data with our internet scan data is based on the data available as of 12 February 2026. According to the providers' own figures, the Lantern dataset comprises around 79,000 FHIR endpoints. However, we could only analyse \LanternAllFHIRHosts ~of these, as the remaining endpoints were not reachable. 

The \LanternAllFHIRHosts ~FHIR endpoints that we analysed are spread across just \LanternFHIRUniqueHosts ~unique hostnames. Of these endpoints, 45,358 contain arbitrary values in the \texttt{\{endpoint\}} section of the URL, so they remain undetected by scans. 
Similarly, some endpoints use a fixed value for the \texttt{\{endpoint\}} part, but utilize different subdomains for the \texttt{\{ip/domain\}} part. 

Nonetheless, the Lantern dataset also contains scannable FHIR URLs, i.e., with fixed endpoint parts. We included the three most commonly used paths in an additional scan: '/v1/uscore/R4/metadata', '/fhir/r4/metadata', and '/fhirproxy/api/fhir/r4/metadata'. 
As a result, we found 659 further endpoints. Of these, 61 endpoints are part of the Lantern dataset, and 416 endpoints belong to systems identified in our initial scan. 
This again demonstrates that some FHIR hosts use multiple endpoints. Ultimately, we found minor overlap between our results and the Lantern dataset. Both datasets include six FHIR endpoints. After adding three endpoints for the scan method, the number of overlaps increased to 61.

Furthermore, we conducted an analysis of the authentication methods used by the endpoints in the Lantern dataset. Our analysis reveals that 28,783 endpoints support the combination of Smart-on-FHIR and OAuth, while 12,246 endpoints support only Smart-on-FHIR. 
A further 354 endpoints support the combination of Basic, Smart-on-FHIR and OAuth. One endpoint supports OAuth exclusively. 
For 10,059 endpoints, it was not possible to determine the authentication method from the metadata. 

With regard to known vulnerabilities, we identified six FHIR endpoints in the Lantern dataset that suffer from at least one CVE-scored vulnerability. Four of these endpoints use an HAPI FHIR server and are affected by CVE-2024-51132. We found two endpoints that use the OpenEMR FHIR server and are affected by 74 vulnerabilities. From transport security perspective, all endpoints except one use TLS. We identified only one system through which patient data could be retrieved. This system contains exactly one patient.

Overall, the analysis of the Lantern dataset shows that the listed FHIR endpoints exhibit a significantly higher level of security than those identified using the scan method. One explanation could be that the FHIR endpoints in the Lantern dataset originate from FHIR developers and professional service providers and are intentionally made publicly accessible. Consequently, the FHIR endpoints have been secured by the operators.

\subsection{Honeypot Portlist}
\label{sec:apendix_honeypot}
The honeypot we have implemented listens on the following ports:

\begin{table}[H]
\centering
\footnotesize
\caption{List of all ports on which the honeypot is listening}
\label{tab:honeyports}
\begin{tabular}{rl}
\toprule
Ports & Protocol \\
\midrule
80 & FHIR \\
443 & FHIR\\
3000 & FHIR \\
3005 & FHIR \\
5000 & FHIR\\
8000 & FHIR \\
8081 & FHIR  \\
8082 & FHIR \\
8083 & FHIR \\
8085 & FHIR \\
8089 & FHIR \\
8090 & FHIR \\
8888 & FHIR \\
9000 & FHIR \\
9080 & FHIR \\
104 & DICOM\\
2761 & DICOM \\
2762 & DICOM \\
11112 & DICOM \\
4242 & DICOM\\
4498 & DICOM \\
2575 & HL7 \\
20046 & HL7 \\
2222 & HL7 \\
\bottomrule
\end{tabular}
\end{table}

\subsection{Disclosure Questionnaire}
\label{sec:apendix_Disclosure}
The following questions were included in the disclosure procedure:

\begin{itemize}
    \item \textbf{Q1:} Were you aware that the system is accessible via the corresponding ports on the Internet? (yes/no)
    \item \textbf{Q2:}  Are the systems intentionally exposed to the Internet? (yes/no)
    \item \textbf{Q3:}  Will you reconfigure the systems so that they are no longer accessible on the Internet in the future? (yes/no)
    \item \textbf{Q4:}  Is contacting you via the abuse email address the best way to inform you of possible security risks, or would you prefer an alternative method?
(text)
\end{itemize}

\end{document}